\documentclass[
aps,
prd,
10pt,
twocolumn,
nofootinbib,
preprintnumbers,
superscriptaddress,
showpacs,
amsmath,
amssymb,
floatfix]{revtex4-2}

\usepackage{comment}
\usepackage{booktabs,makecell}
\usepackage{graphicx}
\usepackage{amsmath,amssymb}
\usepackage{amsfonts}
\usepackage{xspace} 
\usepackage[usenames]{color}
\usepackage{dcolumn}
\usepackage{bm}
\usepackage{mathrsfs}
\usepackage[colorlinks=true,citecolor=blue]{hyperref}
\usepackage[numbered, open=false, openlevel=2]{bookmark}
\usepackage[all]{hypcap} 
\usepackage[utf8]{inputenc} 
\usepackage{slashed}
\usepackage{multirow}
\usepackage{rotating}
\definecolor{orange}{rgb}{1,0.5,0}
\usepackage{tabularx}

\usepackage[normalem]{ulem}

\usepackage[dvipsnames]{xcolor}

\newcommand{\egabb}{e.\,g.\@}
\newcommand{\ie}{i.\,e.\@}
\newcommand{\identifier}[3]{\texttt{{#1}\rlap{\textsubscript{#2}}{\textsuperscript{#3}}}}

\usepackage{natbib}
\usepackage{tabularray}
\usepackage{mathtools}
\usepackage{multirow}
\usepackage{orcidlink}

\definecolor{greenA}{HTML}{10BC10}

\begin{document}
\def\imrphenomxasnrtidalthree{{\texttt{IMRPhenomXAS\_NRTidalv3}}}

\title{Numerical Relativity Simulations of Dark Matter Admixed Binary Neutron Stars}

\newcommand{\affilcoimbra}{CFisUC, Department of Physics, University of Coimbra, Rua Larga P-3004-516, Coimbra, Portugal}
\newcommand{\affilaei}{Max Planck Institute for Gravitational Physics (Albert Einstein Institute), Am Mühlenberg 1, Potsdam 14476, Germany}
\newcommand{\affilpotsdam}{Institut für Physik und Astronomie, Universität Potsdam, Haus 28, Karl-Liebknecht-Str. 24-25, Potsdam, Germany}
\newcommand{\affilcentra}{CENTRA, Departamento de F\'{\i}sica, Instituto Superior T\'ecnico -- IST, Universidade de Lisboa -- UL, Avenida Rovisco Pais 1, 1049-001 Lisboa, Portugal}
\newcommand{\affillondon}{Department of Physics, Royal Holloway University of London, Egham, TW20 0EX}
\newcommand{\affilflorida}{Department of Physics, Florida Atlantic University, Boca Raton, FL 33431, USA}
\newcommand{\affilsoton}{Mathematical Sciences and STAG Research Centre, University of Southampton, Southampton SO17 1BJ, United Kingdom}

\author{Edoardo Giangrandi\,\orcidlink{0000-0001-9545-466X}}\email{edoardo.giangrandi.1@uni-potsdam.de}
\affiliation{\affilcoimbra}
\affiliation{\affilpotsdam}

\author{Hannes R. \surname{Rüter}\,\orcidlink{0000-0002-3442-5360}}
\affiliation{\affilcentra}

\author{Nina Kunert\,\orcidlink{0000-0002-1275-530X}}\email{nkunert@uni-potsdam.de}
\affiliation{\affilpotsdam}

\author{Mattia Emma\,\orcidlink{0000-0001-7943-0262}}\email{mattia.emma@ligo.org}
\affiliation{\affillondon}

\author{Adrian Abac\,\orcidlink{0000-0003-4786-2698}}
\email{adrian.abac@aei.mpg.de}
\affiliation{\affilaei}
\affiliation{\affilpotsdam}

\author{Ananya Adhikari\,\orcidlink{0000-0002-3890-3577}}
\affiliation{\affilflorida}

\author{Tim Dietrich\,\orcidlink{0000-0003-2374-307X}}
\affiliation{\affilpotsdam}
\affiliation{\affilaei}

\author{Violetta Sagun\,\orcidlink{0000-0001-5854-1617}}\email{v.sagun@soton.ac.uk}
\affiliation{\affilsoton}

\author{Wolfgang Tichy\,\orcidlink{0000-0002-8707-754X}}
\affiliation{\affilflorida}

\author{Constança Providência\,\orcidlink{0000-0001-6464-8023}}\email{cp@uc.pt}
\affiliation{\affilcoimbra}

\date{\today}

\begin{abstract}

Binary neutron star mergers provide insight into strong-field gravity and the properties of ultra-dense nuclear matter. These events offer the potential to search for signatures of physics beyond the standard model, including dark matter. We present the first numerical-relativity simulations of binary neutron star mergers admixed with dark matter, based on constraint-solved initial data. Modeling dark matter as a non-interacting fermionic gas, we investigate the impact of varying dark matter fractions and particle masses on the merger dynamics, ejecta mass, post-merger remnant properties, and the emitted gravitational waves. Our simulations suggest that the dark matter morphology - a dense core or a diluted halo - may alter the merger outcome. Scenarios with a dark matter core tend to exhibit a higher probability of prompt collapse, while those with a dark matter halo develop a common envelope, embedding the whole binary. Furthermore, gravitational wave signals from mergers with dark matter halo configurations
exhibit significant deviations from analytical models when the tidal
deformability is calculated in a standard two-fluid framework. This highlights 
the need for refined models in calculating the tidal deformability when
considering mergers with extended dark matter structures. These initial results provide a basis for further exploration of dark matter's role in binary neutron star mergers and their associated gravitational wave emission and can serve as a benchmark for future observations from advanced detectors and multi-messenger astrophysics. 
\end{abstract}

\maketitle

\section{Introduction}

Following the groundbreaking detection of the first binary neutron star (BNS) merger GW170817~\cite{LIGOScientific:2017vwq}, which was accompanied by the detection of its electromagnetic counterparts - the short $\gamma$-ray burst GRB170817A, its afterglow, and the kilonova AT2017gfo~\cite{LIGOScientific:2017pwl,Lipunov:2017dwd,Shappee:2017zly,J-GEM:2017tyx,Goldstein:2017mmi,LIGOScientific:2017zic,LIGOScientific:2017zic} - our understanding of compact stars and their merger dynamics has undergone a significant revolution. This event marked the dawn of gravitational wave (GW) astrophysics for neutron stars (NSs), establishing powerful new tools for probing their internal structure. By combining GW observations with the corresponding electromagnetic signal from the same source, we can now extract information about the properties of matter at its extreme limits~\cite{LIGOScientific:2017vwq, LIGOScientific:2017ync, Cowperthwaite:2017dyu, Read:2009yp, Metzger:2019zeh}. 
The analysis of the GW170817 signal, measured by the advanced LIGO and Virgo
detectors, yielded a constraint on the tidal deformability parameter of the 
$1.4 M_\odot$ NS, $\Lambda_{1.4}\leq800$~\cite{Abbott_2018}. 
Moreover, GW190425~\cite{LIGOScientific:2020aai}, considered a BNS merger candidate, provided constraints on the equation of state (EOS) of NS matter compatible with those obtained from the first event. However, due to a low signal-to-noise ratio, GW190425 did not provide significant new insights into the specific properties of the EOS~\cite{Dudi:2021abi,LIGOScientific:2020aai}.
Complementary to the GW observations, X-ray data from NICER~\cite{Miller:2019cac,Miller:2021qha,Riley:2019yda,Riley:2021pdl,Raaijmakers:2019dks}, 
radio measurements of the heaviest pulsars, 
\ie{}, PSR J0348+0432 with a mass of 
$2.01\pm0.04 M_\odot$~\cite{Antoniadis:2013pzd}
and PSR J0740+6620 with $2.08\pm0.07 M_\odot$~\cite{Fonseca:2021wxt}, optical
observations of the \textit{black widow} pulsars PSR J1810+1744 with 
$2.13\pm0.04 M_\odot$~\cite{Romani:2021xmb} and PSR J0952-0607 with
$2.35\pm 0.17 M_\odot$ \cite{Romani:2022jhd}, thermonuclear accretion bursts 
in low-mass X-ray binaries~\cite{Nattila:2017wtj,Goodwin:2019qbe} along with
nuclear physics terrestrial experiments provide additional constraints on the
properties of NSs and the EOS; see~\cite{Koehn:2024set} for a recent review.
Nevertheless, all the aforementioned analyses and models often employ the
simplifying assumption that NSs reside in a perfect vacuum where no 
dark matter (DM) is present. However, due to the extreme gravitational fields and
compactness, NSs might trap and accumulate DM from the galactic halo within their interiors and the surrounding environments. The presence of DM could have a significant impact on the internal structure~\cite{Panotopoulos:2017idn, Abac:2021txj, Giangrandi:2022wht, Hippert:2022snq,Diedrichs:2023trk}, thermal evolution~\cite{Giangrandi:2024qdb}, and overall compactness of these objects. 
The distribution of DM within NSs depends crucially on its particle properties, such as mass, interaction with the Standard Model particles, 
as well as the amount of trapped DM. Once trapped in the stellar environment, DM may either form a dense core in the inner stellar region or be distributed in a diluted halo that completely embeds the baryonic star~\cite{Ivanytskyi:2019wxd}. In the former case, the increased gravitational pull from the admixed DM leads to more compact NS configurations compared to their purely nucleonic counterparts~\cite{Giangrandi:2024qdb}. On the other hand, a surrounding DM halo can significantly increase the total gravitational mass of the system, pushing the outermost radius of the DM-admixed NS to tens or even hundreds of kilometers~\cite{Ivanytskyi:2019wxd}. In both scenarios, the presence of the additional DM component may deeply modify the matter distribution, affecting the tidal deformability parameter and causing substantial alterations to BNS merger dynamics and the associated GW signature~\cite{Ellis:2017jgp}. The DM fraction accumulated by NSs is highly dependent on their galactic environment. Indeed, NSs located closer to the galactic center or in the DM dense regions are expected to accumulate a greater amount of DM~\cite{Ivanytskyi:2019wxd}. This is due to the significantly higher DM density present in the center of the galaxy~\cite{Stadel:2008pn,Weber:2009pt}. This spatial dependence on DM density adds another layer of complexity to the potential impact of DM on BNS mergers. Moreover, NSs in binary systems have a particularly long evolutionary timescale, allowing for an even greater DM fraction due to the longer accumulation time and the stronger combined gravitational field of the two stars in a binary system. These factors might lead to higher DM fractions in the NSs in a binary compared to isolated stars~\cite{Kouvaris:2010vv}.

Early investigations into the potential DM influence on the GW signals from BNS mergers include pioneering work such as~\citet{Ellis:2017jgp}. They explored the impact of DM cores within NSs on emitted GWs and employed a simplified mechanical model of BNS mergers. Their analysis suggested that DM cores could generate a supplementary peak in the post-merger GW power spectral density, potentially distinguishable from those produced by the Standard Model matter. While providing insights on the possible DM-related features of the GW, the simplified nature of the model may not fully capture the complexity of the merger dynamics, particularly in scenarios involving significant DM fractions, \egabb{}, up to $\sim10\%$, which could need non-standard dark sector models and specific formation scenarios. \citet{Bezares:2019jcb} further investigated GW signatures of BNS mergers with bosonic DM cores using numerical simulations. In their treatment, baryonic matter (BM) was modeled as a perfect fluid and DM as a complex scalar field, with their results indicating the presence of a strong $m=1$ mode in the GW waveform. Such an effect was observed only at high boson-to-fermion fractions of $10\%$. This configuration yielded dark cores causing a mass redistribution, leading to an $|m|=1$ over-density via a one-arm instability. However, the overall dynamical evolution and GW signal remained almost unaffected during the merger and inspiral phases, posing a significant challenge to the identification of DM imprints.  \citet{Bauswein:2020kor} modeled NS binaries with a DM component in their interior, approximating DM as test particles within three-dimensional relativistic simulations. Their simulations suggest that DM remains gravitationally bound within the merger remnant, orbiting inside the BM structure and producing GW signals in the kHz range.  \citet{Emma:2022xjs} investigated the merger of BNS systems that contain mirror DM.  By performing single-star tests and BNS simulations, their findings suggest that the presence of DM reduces the lifetime of the merger remnant, favoring a prompt collapse to a BH. However the initial data (ID) were constructed using superimposed boosted single-star spacetimes, a simplification that introduces limitations in accurately modeling realistic binary systems. This issue has been addressed by~\citet{Ruter:2023uzc}, where the \textsc{sgrid} code~\cite{Tichy:2009yr,Tichy:2012rp,Dietrich:2015pxa,Tichy:2019ouu} was updated to construct quasi-equilibrium configurations for DM-admixed stars, focusing on the ID problem. \citet{Owen:2025odr} explored BNS mergers with DM interacting via a massive dark abelian vector field. They calculated modifications to the binary inspiral waveform to the first post-Newtonian order, adding such corrections into a state-of-the-art waveform model.
\citet{Suarez-Fontanella:2024epb} investigated the dynamics in the presence of a viscous DM-polluted environment, using an effective Lagrangian model. The model qualitatively reproduces the GW signal observed in early post-merger phases.
While non-viscous DM shows negligible impact on the power spectral density up to a DM fraction of 5\%, the inclusion of DM leads to significant damping of the main peak without any frequency shifts.  Furthermore, the analysis suggests that this damping could make it difficult for current interferometers to observe key spectral features, which might be mistaken for a prompt collapse.  These effects could have implications for the detection and interpretation of GW signals by current and future interferometers, such as LIGO-Virgo-KAGRA, the Einstein Telescope~\cite{Branchesi:2023mws, Punturo:2010zz, Tichy:2009yr,Abac:2025saz}, Cosmic Explorer~\cite{Reitze:2019dyk, Reitze:2019iox} and NEMO~\cite{Ackley:2020atn}.  To further quantify the DM influence on the BNS coalescence, we present in this work, up to our knowledge, the first consistent simulations using constraint-solved ID of DM-admixed BNS systems.  Moreover, we utilize tabulated microphysical EOSs for both fluids, offering a more detailed and realistic model for each component.  Another key feature of our simulations is the inclusion of all DM morphologies, including DM halos embedding the stars, which further distinguishes our work from previous studies~\cite{Emma:2022xjs}.

The present article is organized as follows. In Section~\ref{section:setup}, 
we discuss the \textsc{sgrid} code for the ID construction and the 
\textsc{BAM} code for the time evolution.
Section~\ref{section:results} shows the results of this study. We begin by
presenting the ID configurations in~(\ref{subsection:ID}), followed by 
the analysis of the time evolution in~(\ref{subsection:evolution}), 
the fate of the remnant~(\ref{subsection:remnantproperties}), the angular-velocity evolution~(\ref{subsection:Angular-velocity}), the ejecta masses~(\ref{subsection:ejecta}), and the extracted GWs~(\ref{subsection:GW}). Conclusions are presented in Section~\ref{section:conclusions}. In Appendix~\ref{AppendixA:L2norm}, we present the time evolution of the L2-norm followed by a discussion of GW convergence in Appendix~\ref{AppendixA:GWconvergence}.

\section{Setup}\label{section:setup}

\subsection{Framework}
In this study, we model DM within NSs as a non-interacting fermionic fluid, coupled only through gravity with BM. While acknowledging the possible existence of non-gravitational interactions, we justify this simplification through a combination of observational and experimental constraints.
This type of DM model can serve as a simplified framework for exploring GeV to sub-GeV fermionic weakly interacting massive particles, which could describe WIMPs~\cite{Roszkowski:2017nbc}, asymmetric DM~\cite{Giangrandi:2022wht}, hidden sector DM~\cite{Pospelov:2007mp} or mirror DM scenarios~\cite{Foot:2014mia}.
The observed separation of DM and BM distributions in galaxy clusters collisions, \egabb{}, the Bullet Cluster~\cite{Paraficz:2012tv,Robertson:2016xjh}, strongly suggests that DM and BM feebly interact.
The observed spatial offset would be significantly smaller if DM and BM had significant non-gravitational interaction. Moreover, direct detection experiments provide an upper limit on the DM-nucleon scattering cross-section which is many orders of magnitude lower than the typical nuclear one~\cite{PhysRevLett.131.041002,XENON:2023cxc}. Based on these constraints and considering the very short timescales of the merger, the 
approximation of no interaction between DM and BM, except
through gravity, is fully justified. Therefore, for non-interacting fluids, the energy-momentum tensor can be split into two individual components~\cite{PhysRevD.102.063028,Rafiei_Karkevandi_2022,Emma:2022xjs}, as
\begin{equation}\label{eq:SET-2F-perfect}
    T_{\mu\nu}^{(s)} = (\varepsilon^{(s)}+p^{(s)})u_\mu^{(s)} u_\nu^{(s)} + p^{(s)} g_{\mu\nu}\,,
\end{equation}
where the index $s\in\{\mathrm{BM},\mathrm{DM}\}$ labels the fluid component, $\varepsilon$, $p$, and $u^\mu$ are the energy density, pressure, and the four-velocity of the fluid, respectively.

The Einstein field equations are given by
\begin{equation}
    R_{\mu\nu}+\frac{1}{2}g_{\mu\nu}R = 8\pi \sum_s T_{\mu\nu}^{(s)}\,.
\end{equation}
Additionally, each of the components satisfies the energy-momentum conservation separately, namely:
\begin{equation}
    \nabla^\mu T_{\mu\nu}^{(s)} = 0\,.
\end{equation}
For each fluid, we can also define the specific enthalpy $h^{(s)}$
\begin{equation}
 h^{(s)} = \frac{\varepsilon^{(s)} + p^{(s)}}{\rho^{(s)}}\,,
\end{equation}
being computed using the fluid rest-mass density $\rho^{(s)}$.

\subsection{Initial data}\label{subsubsec:ID}
The initial configurations for the BNS simulations in this work are constructed
using the pseudo-spectral code \textsc{sgrid}~\cite{Tichy:2009yr,Tichy:2012rp,Dietrich:2015pxa,Tichy:2019ouu}.

In the construction of the initial data we control our desired target masses
for the individual fluid components.
Since in a binary system
the gravitational mass of the individual stars is not well defined, we
instead control the rest masses $M_0^{(s)}$ of the stars.
For a binary system we then define the gravitational component masses $M^{(s)}$ 
to be that of the corresponding Tolman-Oppenheimer-Volkoff (TOV)-like 
DM-admixed NS with these rest masses.
The gravitational mass of an individual fluid component is computed following Eq.~(8) of Ref.~\cite{Giangrandi:2022wht}.
We define the DM fraction in terms of these gravitational component masses by
\begin{equation}
    f_\mathrm{DM} := \frac{M^{(\mathrm{DM})}}{M^{(\mathrm{BM})}+M^{(\mathrm{DM})}}\,.
\end{equation}

The \textsc{sgrid} code employs surface fitting coordinates to solve the Einstein constraint equation through the extended conformal thin sandwich (XCTS) formalism~\cite{York:1998hy,Pfeiffer:2002iy}. Using the adapted version of \textsc{sgrid} presented in Ref.~\cite{Ruter:2023uzc}, DM-admixed BNS configurations are constructed and used as ID for our simulations.
The \textsc{sgrid} code deals with the spacetime metric $g_{\mu\nu}$ using the commonly used $3+1$ decomposition, where the line element can be written as follows
\begin{equation}
    ds^2 = -\alpha dt^2 + \gamma_{ij}(\beta^i dt +dx^i)(\beta^j dt +dx^j)\,,
\end{equation}
with $\alpha$, $\beta^i$, and $\gamma_{ij}$ being the lapse, shift, and the spatial part of the metric tensor $g_{\mu\nu}$ induced on three-dimensional spatial hypersurfaces, respectively. The latter can be written down as
\begin{equation}
    \gamma_{\mu\nu}=g_{\mu\nu}+n_{\mu}n_{\nu}\,,
\end{equation}
with $n^\mu$ being the timelike normal vector to the three-dimensional hypersurfaces defined by
\begin{equation}
  \label{eq:timelike_normal}
    n^{\mu}=\frac{1}{\alpha}(1,-\beta^i)\,,\quad {\rm and}\quad n_{\mu}=(-\alpha, 0,0,0)\,.
\end{equation}
To generate the initial configurations we assume that the system is in
quasi-equilibrium at vanishing temperature, $T=0$.
We first solve the equations for the velocity potentials $\phi^{(s)}$, which are defined through the following four-velocity
\begin{equation}\label{eq:4velEq}
    \gamma^i_\mu u^{(s)\mu} = \frac{1}{h^{(s)}}(D^{i} \phi^{(s)} + w^{(s) i})\,,
\end{equation}
with $w^{(s) i}$ being a divergence-free vector, representing the rotational part of the fluid component. Following the same approach shown in~\cite{Ruter:2023uzc}, we can rewrite the final equations derived from Eq.~\eqref{eq:4velEq} as
\begin{equation}\label{eq:velpotEq}
    D_i \left( \frac{\alpha\rho^{(s)}}{h^{(s)}} (D^{i} \phi^{(s)} 
    + w^{(s) i}) -\alpha\rho^{(s)} u^{(s)0} (\beta^i+\xi^i) \right) = 0\,, 
\end{equation}
where the approximate Killing vector $\xi$ was introduced in order to set the time derivatives of the fields.

To facilitate the convergence of the solver to a physically realistic solution, 
we construct our initial guess by superimposing two boosted 
TOV-like DM-admixed NSs, each with our desired target gravitational masses.
Generating TOV-like solutions with the target masses requires determining 
the central conditions for both fluid components, which is a two-dimensional 
root-finding problem. This can be easily achieved through an iterative bisection method applied to the central pressure of one fluid while holding the central pressure of the other fluid constant. This process allows us to accurately determine the necessary central pressures to achieve the desired total mass of the two-fluid structure. More advanced iterative algorithms, such as the Newton-Raphson, present challenges as the method may converge to a local extremum of the mass function, effectively halting the iterations. 

We start by evaluating the residuals of Eq.~\eqref{eq:velpotEq}.
If the residuals are above $10\%$ of the combined residuals of the 
XCTS equations~\cite{2010nure.book.....B,Tichy:2016vmv} we solve for the
velocity potentials in Eq.~\eqref{eq:velpotEq}.
Subsequently, we update the four-velocity by taking a weighted average of the previous solution $\phi_\mathrm{old}^{(s)}$ and the newly obtained one $\phi_\mathrm{new}^{(s)}$ as
\begin{equation}
    \phi^{(s)} = w \phi_\mathrm{old}^{(s)} + (1-w)\phi_\mathrm{new}^{(s)}\,.
\end{equation}
In this iterative process, $w$ represents a weighting factor designed to enhance
the numerical stability of the iteration.
To achieve a suitable balance between considering new information and preserving numerical stability, we set $w=0.85$, reducing the chances of overshooting during the iteration process~\cite{Moldenhauer:2014yaa}.
Subsequently we solve the XCTS equations, updating the values of 
$\alpha$, $\beta$, and the conformal factor $\psi$, using the same weighted 
average of old and new solutions.

At the end of each iteration we adjust free parameters governing the specific
enthalpy $h^{(s)}$ to obtain the correct target rest-masses for both fluids,
in the same way as described in section~II of~\cite{Ruter:2023uzc}. 
If the sum of the residuals at this stage is larger than a user-chosen threshold 
and the number of iterations is below a prescribed limit, the algorithm 
repeats the procedure from the point of evaluating the residuals
of Eq.~\eqref{eq:velpotEq}.
Otherwise the iteration stops and \textsc{sgrid} performs a final solving
of the XCTS equations.

To close the system, EOSs are needed, either as the parameters of piecewise polytropes or as full tabulated EOSs. The tabulated EOSs are then interpolated in \textsc{sgrid} in a thermodynamically consistent way using a cubic Hermite interpolation. The considered BM and DM EOSs are discussed in Section~\ref{eossubsection}.

Another important aspect of the \textsc{sgrid} code is that it uses surface-fitting coordinates. This approach is designed to avoid (or reduce) numerical Runge oscillations~\cite{Runge1901a}. These oscillations arise when using high-degree polynomial interpolations to approximate functions describing the star's physical properties, such as density or pressure. This phenomenon leads to physical inaccuracies, particularly near the surface of the star, where the solution is not smooth. With each update to the specific enthalpy $h^{(s)}$, the computational grid is adapted to ensure that the boundaries of the spectral elements align with the new outer fluid surface. 
In our current implementation, surface fitting is adapted to binary
systems and does not allow fluid surfaces that overlap, 
\egabb{}, DM-halo common envelope.
Furthermore, we do not construct domains that are fitted to the surface
of the inner fluid and hence the Runge phenomenon is expected to be observed at the surface of the inner fluid. Extending the code's capabilities to handle more complex scenarios, such as a DM common envelope in the ID, would necessitate a substantial restructuring of the computational framework. 

\subsection{Dynamical Evolution}
The \textsc{BAM} code is used for the dynamical evolution of the matter and spacetime fields~\cite{Bruegmann:2006ulg,Thierfelder:2011yi,Dietrich:2015iva,Bernuzzi:2016pie,Dietrich:2018phi,Neuweiler:2024jae,Gieg:2024jxs,Schianchi:2023uky}. The code solves Einstein's field equations exploiting the $3+1$ decomposition, cf. Section~\ref{subsubsec:ID}. In the presented runs, the Z4c formulation of the field equations of General Relativity with constraint damping terms~\cite{Bona:2003fj,Bona:2005pp,Gundlach:2005eh,Bernuzzi:2009ex, Hilditch:2012fp}, coupled with the moving puncture gauge $1+\log$-slicing and gamma-driver shift conditions~\cite{Alcubierre:2002kk}, was employed for the spacetime evolution. 

This work focuses on ideal general-relativistic hydrodynamics (GRHD) simulations, excluding magnetic fields and neutrino interaction. In our two-fluid approach to model DM and BM within the numerical simulations, we represent each component as a separate fluid that exists in the same space domain. Within this framework, matter variables are evolved using the Valencia formulation~\cite{1997ApJ...476..221B}. The evolution equations for ideal GRHD are derived from the conservation laws of particle number and energy-momentum:
\begin{equation}
    \nabla_\mu(\rho^{(s)} u^{(s)\mu})=0\,,\quad{\rm and}\quad \nabla_\mu T^{(s)\mu\nu}=0\,,
\end{equation}
with $\nabla_\mu$ being the covariant derivative. To write the evolution system in the form of a balance law as
\begin{equation}\label{eq:blaw}
    \frac{\partial {\bf q}^{(s)}}{\partial t}+\frac{\partial {\bf F}^{(s)i}}{\partial x^i}={\bf s}^{(s)}\,,
\end{equation}
we define the primitive hydrodynamic variables, ${\bf w}^{(s)}=[p^{(s)},\rho^{(s)}, \varepsilon^{(s)}, v^{(s)i}]$, and the conservative variables ${\bf q^{(s)}}=[D^{(s)},S_i^{(s)},\tau^{(s)}]$:
\begin{align}
    D^{(s)}     & = \sqrt{\gamma}\rho^{(s)}W^{(s)}\,,\\
    S_i^{(s)}   & = \sqrt{\gamma} W^{(s)2}h^{(s)}\rho^{(s)}v^{(s)}_i\,,\\
    \tau^{(s)}  & = \sqrt{\gamma}(W^{(s)2}h^{(s)}\rho^{(s)}-p^{(s)})-D^{(s)}\,,
\end{align}
where $\gamma$ is the determinant of the spatial metric $\gamma_{ij}$ of the $3+1$ decomposition of the spacetime and $W^{(s)}$ is the Lorentz factor relative to the Eulerian observer. 
We can obtain the evolution equations in the form of Eq.~\eqref{eq:blaw}:
\begin{equation}
    {\bf q}^{(s)}=[D^{(s)},S_i^{(s)},\tau^{(s)}]\,,
\end{equation}
\begin{equation}
    {\bf F}^{(s)i}=\alpha \left[ \begin{array}{c} D^{(s)}\tilde{v}^{(s)i} \\
    S_j^{(s)}\tilde{v}^{(s)i} + \sqrt{\gamma } p^{(s)} \delta^i_j \\
    \tau^{(s)}\tilde{v}^{(s)i}+ \sqrt{\gamma} p^{(s)} v^{(s)i} \end{array} \right]\,,
\end{equation}
\begin{equation}
    {\bf s}^{(s)} = \alpha \sqrt{\gamma} \left[ \begin{array}{c} 0 \\ 
    T^{(s)\mu\nu} \left( \frac{\partial g_{\nu j}}{\partial x^\mu} - \Gamma^\lambda_{\mu\nu} g_{\lambda j} \right) \\ 
    \alpha \left( T^{(s)\mu 0} \frac{\partial \ln(\alpha)}{\partial x^\mu} - T^{(s)\mu\nu} \Gamma^0_{\mu\nu} \right) \end{array} \right]\,,
\end{equation}
where $\tilde{v}^{(s)i} = v^{(s)i}-\beta^i/\alpha$ and $\Gamma^\lambda_{\mu\nu}$ are the Christoffel symbols.

\subsection{BM and DM EOSs}\label{eossubsection}
For both components, we employ zero-temperature tabulated EOSs. 
BM is modeled with the SLy4 EOS~\cite{Gulminelli:2015csa}, which supports
isolated NSs with the maximum gravitational mass of $2.06  M_\odot$, 
with a radius $R_{\mathrm{M}_{\rm max}}=10.02$ km. We note that SLy4 EOS, being relatively soft, is able to support NSs with tidal deformabilities  $\tilde{\Lambda}\sim400$ for a $1.35 M_\odot$ NS, consistent with current observations and constraints~\cite{LIGOScientific:2018cki}.  The SLy4 EOS is selected due to its established and widespread use within the NR community. BM thermal effects are taken into account as prescribed in~\cite{Bauswein:2010dn, Zwerger:1997sq} with the thermal contribution to the total pressure modeled as $p^{(s)}_\mathrm{th}=(\Gamma^{(s)}_\mathrm{th}-1)\rho^{(s)}\varepsilon^{(s)}_\mathrm{th}$. 
In our simulations, we use a thermal adiabatic index 
$\Gamma^{(\mathrm{BM})}_\mathrm{th}=1.75$. 

DM is modeled as a non-interacting fermionic gas of massive particles with spin
$1/2$ and particle mass $m_\mathrm{DM}$~\cite{Ivanytskyi:2019wxd, Sagun:2021oml, Ruter:2023uzc}.
DM is treated as isothermal, with a thermal adiabatic index of 
$\Gamma_\mathrm{th}^{(\mathrm{DM})}=1$, resulting in $p^{(s)}_{\mathrm{th}}=0$, so that thermal 
contributions of DM to the pressure are neglected. Moreover, to analyze 
the effect of the morphology of the DM structure on the dynamics of the 
merger, we select two particle masses for our simulations: $1\ \rm GeV$ and
$0.17\ \rm GeV$. These values are consistent with those extensively discussed in the literature for producing different DM halos and cores~\cite{Ivanytskyi:2019wxd, Sagun:2021oml, Ruter:2023uzc}.

\subsection{Atmosphere}
In grid-based numerical relativity (NR) simulations, an artificial atmosphere is typically introduced in the vacuum region surrounding compact objects. This is due to the fact that extremely low rest-mass densities pose a significant numerical challenge, potentially leading to inaccuracies in the recovery of the primitive variables and within high-resolution shock-capturing schemes~\cite{Rezzolla:2013dea,Baiotti:2016qnr}. Our artificial atmosphere is
implemented as a cold static fluid.  The density within the atmosphere for each fluid is set to a fraction $f_\mathrm{atm}$ of the initial maximum density as $\rho_\mathrm{atm}^{(s)}= f_\mathrm{atm} \max\limits_{\mathbb{R}^3} \rho^{(s)}(t=0)$.  
The atmosphere pressure and internal
energy are then determined using the zero-temperature part of each EOS, while the fluid velocity is set to zero. To prevent fluctuations near the density floor, we define a threshold density, $\rho_\mathrm{th}^{(s)}=f_\mathrm{th} \rho_\mathrm{atm}^{(s)}$, in terms of the fraction $f_\mathrm{th}$ of the atmospheric density~\cite{Thierfelder:2011yi}. 
Within the \textsc{BAM} code, we set a grid cell to atmosphere when its density falls below such a threshold.  For the BNS runs presented in
Section~\ref{section:results}, we use $f_\mathrm{atm}=10^{-11}$ and 
$f_\mathrm{th}=10$ for the BM and DM components.

\subsection{Grid setup}

The \textsc{BAM} code employs cell-centered nested grids with $L$ refinement
levels, indexed by $\ell=0, ..., L-1$. Each level contains one or more boxes defined by a constant grid spacing $h_\ell$ and $n$ points per direction. A 2-to-1 refinement strategy ensures that the resolution in each level is given by $h_\ell=h_0/2^\ell$. Finer levels with $\ell \geq \ell_\mathrm{move}$ move dynamically, tracking the motion of the NSs or black holes (BHs)~\cite{Bruegmann:2006ulg,Thierfelder:2011yi,Dietrich:2015iva,Dietrich:2018bvi}. 
In this work, we used $\ell=7$ or $\ell=8$ for DM-core (halo) configurations and $\ell_\mathrm{move}=2$. 
This choice was motivated by the necessity of fully enclosing the stars within the finest grid level, especially for the DM-halo configurations, which have significantly larger outermost radii compared to the core configurations.
Thus, in all cases, the outer fluid, either DM (for DM-halo configurations) or BM (for DM-core configurations), was entirely contained within the highest refinement level.
This ensured a reliable resolution for a meaningful comparison between the two scenarios. We simulated three distinct resolutions ${\tt R1}$, ${\tt R2}$, and ${\tt R3}$, corresponding to 96, 144, and 192 grid points per direction at the
finest level, respectively. The non-moving grids have the following number of 
grid points in each direction: ${\tt R1}_{\rm NM}=128$, ${\tt R2}_{\rm NM}=192$
and ${\tt R3}_{\rm NM}=256$. The levels consist of two boxes that individually enclose each star. For all moving levels, boxes are created around each object; when these boxes overlap, a single bounding box is formed. In this work, we employ the method of lines with a fourth-order Runge-Kutta scheme and a Berger-Oliger algorithm for the refinement levels for the time evolution~\cite{BERGER198964, BERGER1984484}. The spacetime part utilizes a finite difference scheme with centered fourth-order stencils to compute spatial derivatives. Hydrodynamical variables are represented by a finite volume formalism, incorporating high-resolution shock-capturing schemes to calculate numerical fluxes between cells. We use WENOZ~\cite{2008JCoPh.227.3191B} reconstruction of characteristic fields together with the LLF Riemann solver. Furthermore, the conservative adaptive mesh refinement strategy, as outlined in~\cite{Dietrich:2015iva}, was used for both fluids to ensure the conservation of baryonic mass, energy, and momentum.

\section{Results}\label{section:results}
This section presents the main results of our NR simulations exploring the properties of non-interacting fermionic DM-admixed BNS mergers.
We first give an overview over the initial data of the simulated configurations~(Section~\ref{subsection:ID}) and continue with the
analysis of the key features in the time evolution of these systems (Section~\ref{subsection:evolution}). We follow-up with
a more in-depth analysis of the 
angular-velocity profiles (Section~\ref{subsection:Angular-velocity}), 
remnant properties (Section~\ref{subsection:remnantproperties}), 
ejecta masses (Section~\ref{subsection:ejecta}), 
and the extracted GWs (Section~\ref{subsection:GW}). 

\subsection{Initial Data}\label{subsection:ID}
We run a total of six different DM-admixed BNS configurations, summarized in
Table~\ref{table:IDs}. To better study how the DM impacts the remnant fate,
we choose two different total masses $2.8 M_\odot$ and $2.4 M_\odot$, 
but restrict to equal mass and irrotational setups. We identify these setups
as \identifier{M28}{}{} and \identifier{M24}{}{} in the text, respectively.
Regarding the DM content, we choose three different DM fractions, 
namely $0\%$ (purely baryonic), $0.5\%$ for DM halos, and $3\%$ for DM cores.
These choices ensure that the outer fluid components of the 
DM-admixed NSs do not overlap, avoiding the formation of a common DM 
envelope in the ID. We note that the chosen DM fractions represent 
a significant amount of DM and are intentionally selected as an extreme 
case to highlight potential DM effects in these scenarios.
All ID are constructed using the updated \textsc{sgrid} code, employing the two-fluid formalism. For purely baryonic configurations we specifically set the DM component to zero. Achieving physically realistic and acceptably low eccentricities, such as $\epsilon\lesssim10^{-3}$, proves difficult in these 
DM-free cases.  This issue arises from limitations within the implementation
of two-fluid framework within \textsc{sgrid} that lacks a stabilizing 
mechanism that prevents the stars from drifting apart during each iteration.
However, the exact source of these difficulties requires further investigation.
This problem is exacerbated when the stars are close to each other. 
For ID affected by this problem, the stars plunge towards each other, 
when evolved with the \textsc{BAM} code, thus hindering our ability to 
reduce the orbital eccentricity.
Since the problematic configuration is purely baryonic, a solution would be to
resort to \textsc{sgrid}'s single fluid solver, but to ensure absolute 
consistency between different configurations we chose to construct its ID with
the two fluid solver as well.
Instead, we opted for a larger initial separation $d_{\rm in}$ in the DM-free simulations, which also ensures stable and physically meaningful initial
conditions.

Furthermore, Table~\ref{table:IDs} lists the tidal deformability of 
corresponding isolated DM admixed NSs with the same rest masses as the stars 
in the ID, 
following the standard prescription outlined 
in~\cite{Das:2020ecp,Giangrandi:2022wht}. Within our two-fluid formalism, the tidal deformability $\Lambda$ is computed by integrating Love's ordinary differential equation~\cite{Hinderer_2008}
up to the outermost radius $R_{\rm out}$, which is defined as the BM radius (for DM-core configurations) or the DM one (for DM-halo configurations).

\begin{table*}[t]
\begin{tabular}{lcccccccccc}
\toprule
 identifier & $m_\mathrm{DM}$ {[}GeV{]} & $f_\mathrm{DM}$ {[}\%{]} & $2M_\mathrm{TOV}~[M_\odot]$&
 $M_\mathrm{ADM}~[M_\odot]$ & $J_\mathrm{ADM}~[M_\odot^2]$ &
 $R^{(\mathrm{BM})}$ [km] & $R^{(\mathrm{DM})}$ [km] & $d_{\rm in}$ [km] &
 $\Lambda^{\rm out}$ & DM Morphology \\
\midrule
\identifier{M24}{00}{~~} & - & 0 & 2.40 & 2.382 & 6.39 & 11.400 & - & 53.05 & 818 & None \\
\identifier{M24}{3C}{~~} & 1 & 3 & 2.40 & 2.381 & 6.08 & 11.154 & 5.329 & 47.02 &  730 & Core \\
\identifier{M24}{05H}{~~~} & 0.17 & 0.5 & 2.40 & 2.381 & 6.09 & 11.377 & 18.645 & 46.91 & 2908 & Halo\\
\midrule
\identifier{M28}{00}{~~} & - & 0 & 2.80 & 2.778 & 8.38 & 11.407 & - & 56.02 & 310 & None \\
\identifier{M28}{3C}{~~} & 1 & 3 & 2.80 & 2.774 & 7.91 & 11.143 & 5.122 & 47.07 & 234 &  Core \\
\identifier{M28}{05H}{~~~} & 0.17 & 0.5 & 2.80 & 2.774 & 7.86 & 11.379 & 16.575 & 46.98 & 901 & Halo  \\
\bottomrule
\end{tabular}
\caption{Overview of simulation parameters and configuration of the ID. From left to right: simulation
identifier, DM particle mass $m_\mathrm{DM}$, DM mass fraction $f_\mathrm{DM}$,
twice the gravitational mass 
$M_\mathrm{TOV} = M^{(\mathrm{BM})}+M^{(\mathrm{DM})}$ 
of corresponding isolated TOV-like NSs,
ADM mass $M_\mathrm{ADM}$, 
ADM angular momentum $J_\mathrm{ADM}$,
BM radius $R^{(\mathrm{BM})}$ in canonical Schwarzschild coordinates, 
DM radius $R^{(\mathrm{DM})}$ in canonical Schwarzschild coordinates, 
the initial orbital separation $d_{\rm in}$ in the adapted coordinates of the 
XCTS system and the tidal deformability $\Lambda^{\rm out}$ calculated using 
the outermost radius following Ref.~\cite{Leung:2022wcf,Giangrandi:2022wht}, and DM morphology, respectively.
}
\label{table:IDs}
\end{table*}

\subsection{Evolution}\label{subsection:evolution}
\begin{figure}[t]
    \includegraphics[width=\linewidth]{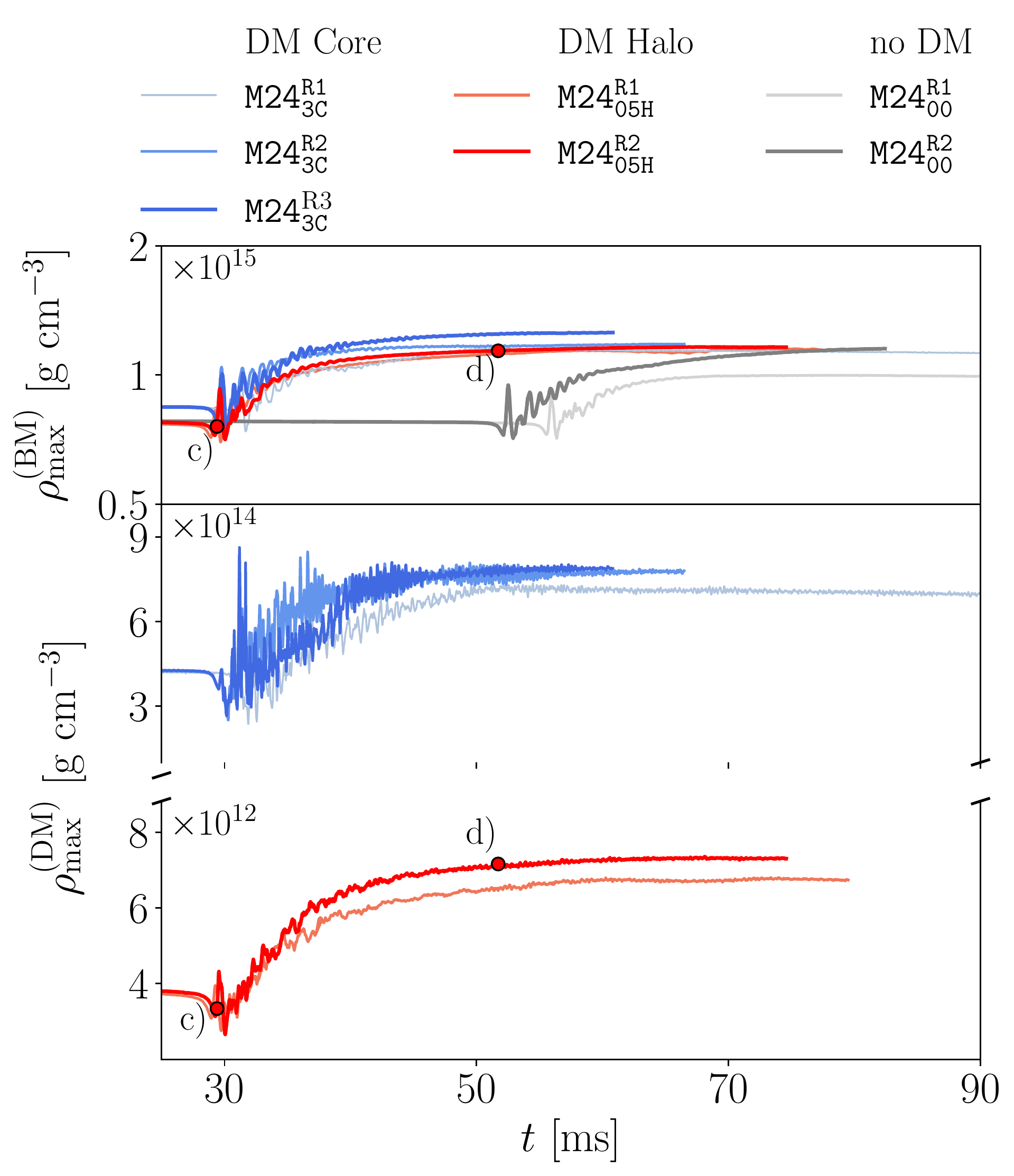}
    \caption{Comparison of the central rest-mass densities during the merger 
    and post-merger phase for BM (upper panel) and DM (lower panel) for the
    systems with the total gravitational mass of $2.4~ M_\odot$. 
    The central rest-mass density is extracted from the finest refinement 
    level in each run. The points labeled with letters indicate the specific 
    times that correspond to the 2D slices shown in Fig.~\ref{fig:panel_halo}.
    }
\label{fig:rho_max_evolution_24}
\end{figure}

Figure~\ref{fig:rho_max_evolution_24} shows the evolution of the central 
rest-mass density of BM and DM for the \identifier{M24}{}{} simulations 
presented here. Before the merger, defined as the point at which the GW 
amplitude reaches its peak value, we observed minor initial transient density
oscillations in both fluids. These transients could arise from the 
interpolation of the ID, from the approximate choice of the free data in the XCTS equations, and from the absence of surface-fitting coordinates within \textsc{sgrid} for the inner fluid.
Right before the merger, the central densities of both the BM and DM components experience a slight decrease as the binary separation shrinks. This effect is due to the increasing tidal forces, which stretch the stars, resulting in a lower maximum density. All simulations reach a stable post-merger configuration. In simulations with DM cores, the presence of DM within the remnant leads to an increased gravitational pull compared to the DM-free runs. This yields more compact remnants with higher central BM densities. On the other hand, while DM halos still impact the overall merger dynamics, their gravitational effect on the central region is less pronounced. The halo contributes to the total gravitational mass of the system, affecting the orbital evolution and the merger time. However, it does not significantly affect the matter distribution of the remnant, showing that the overall effect on BM is minimal. 
\begin{figure*}[ht]
\includegraphics[width=0.95\textwidth]{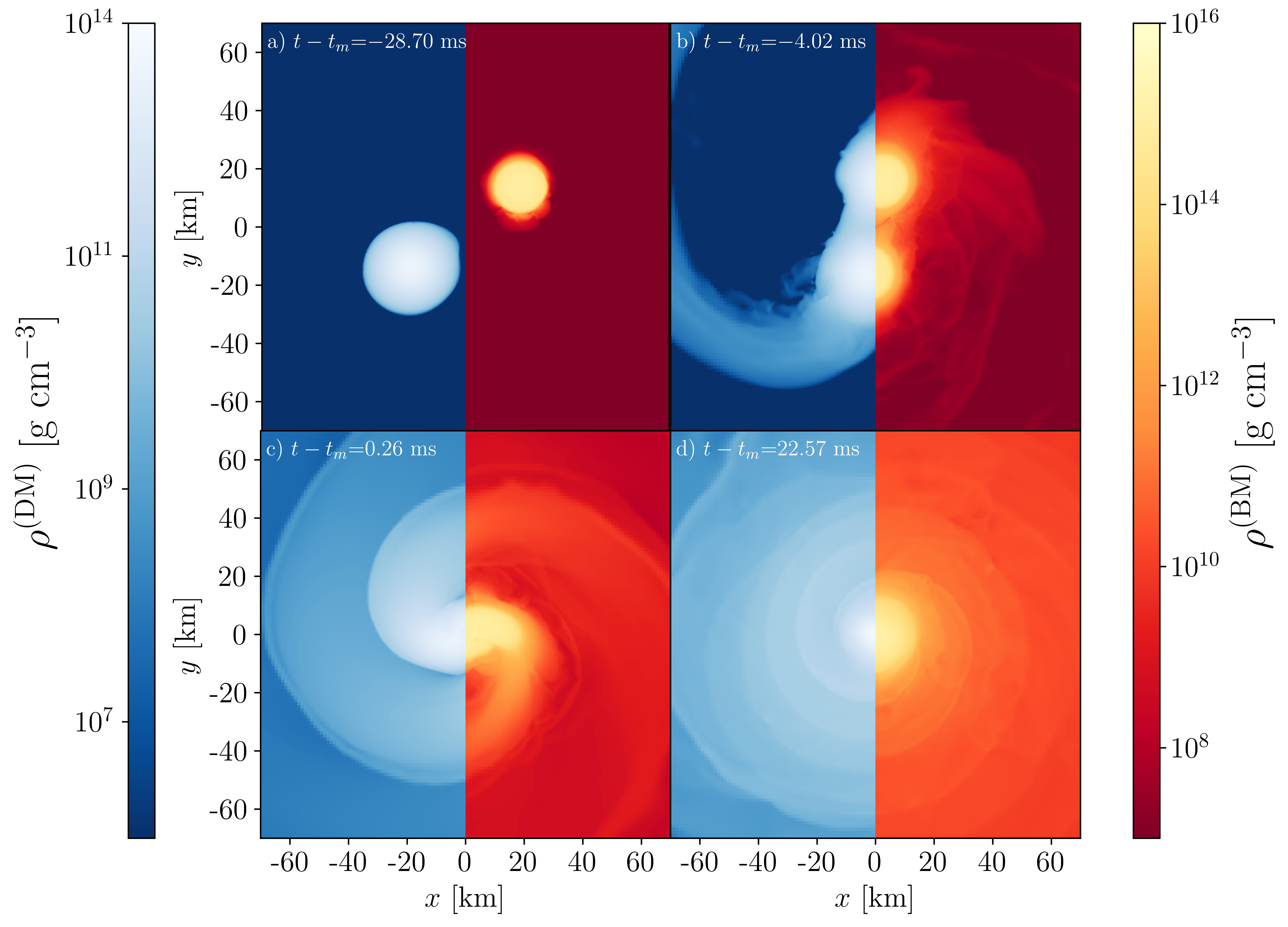}
    \caption{Equatorial density distributions for both BM and DM components, for ${\tt M24}_{\tt 05H}^{\tt R2}$ run at selected times for $M_\mathrm{tot}=2.4~ M_\odot$ with $f_\mathrm{DM}=0.5\%$. The presented times are $t-t_m=(-28.7, -4.02, 0.26, 22.57)$ ms for the upper left, upper right, lower left, and lower right panel, respectively. Times c) and d) are the ones marked in Fig.~\ref{fig:rho_max_evolution_24} with the corresponding labels. The left halves (negative $x$) show the DM density, whereas the right halves (positive $x$) show the BM density.} 
    \label{fig:panel_halo}
\end{figure*}

To further examine the dynamics and better visualize the merger, 
in Fig.~\ref{fig:panel_halo} we show the distribution of the BM and DM
rest mass density on spatial slices through the orbital plane
at four snapshots of the \identifier{M24}{R2~}{05H}~simulation.
During the early inspiral phase shown in panel a),
at $t-t_m=-28.7$ ms ($t_m$ being the merger time),
the DM clearly extends beyond the BM components, forming dilute DM halos.
The initial central BM density within each star is 
$\rho_c^{(\rm BM)}=7.837\cdot10^{14}\, \mathrm{g}/\mathrm{cm}^{3}$, 
while the initial central DM-halo density is 
$\rho_c^{(\rm DM)}=3.910\cdot10^{12} \, \mathrm{g}/\mathrm{cm}^{3}$, \ie{}, two orders of magnitude smaller.
As the system evolves, tidal forces become more dominant. 
While BM just begins to show signs of deformation, the DM halos also respond to the gravitational influence of the companion star, developing a DM bridge with $\rho^{\rm (DM)}\sim 10^{9} \, \mathrm{g}/\mathrm{cm}^{3}$ between the two stellar structures. At $t-t_m=-4.02$ ms, panel b) of Figure~\ref{fig:panel_halo}, the DM components of the two NSs have already begun to merge, forming a common DM envelope that embeds both baryonic cores. This common DM envelope has implications for the subsequent evolution of the system, affecting the dynamics of the merger of the BM component and the properties of the final remnant. 
The BM distributions are also highly deformed, with a clear indication of a
developing BM bridge connecting the two stars at densities around 
$10^{10}\, \mathrm{g}/\mathrm{cm}^{3}$. 
Two key time points c) and d) are marked in Figure~\ref{fig:rho_max_evolution_24}
and correspond to the bottom panels of Figure~\ref{fig:panel_halo}. 
These points represent different stages of the merger: c) the merger time, 
and d) the post-merger phase, showing the hypermassive NS (HMNS) remnant.
The post-merger snapshot at $t-t_m=+22.57$ ms reveals the formation of a complex remnant.
The BM has already merged into a central core with a maximum BM density of 
$\rho_\mathrm{max}^{\rm (BM)}=1.138\cdot10^{15}\, \mathrm{g}/\mathrm{cm}^{3}$,
surrounded by a more diffuse DM component, with a maximum DM density of 
$\rho_\mathrm{max}^{\rm (DM)}=7.117\cdot10^{12}\, \mathrm{g}/\mathrm{cm}^{3}$.
However, unlike the DM-core scenario, the DM distribution still shows an extended halo-like morphology, albeit with a higher central density. 

\begin{figure}[t]
    \includegraphics[width=\linewidth]{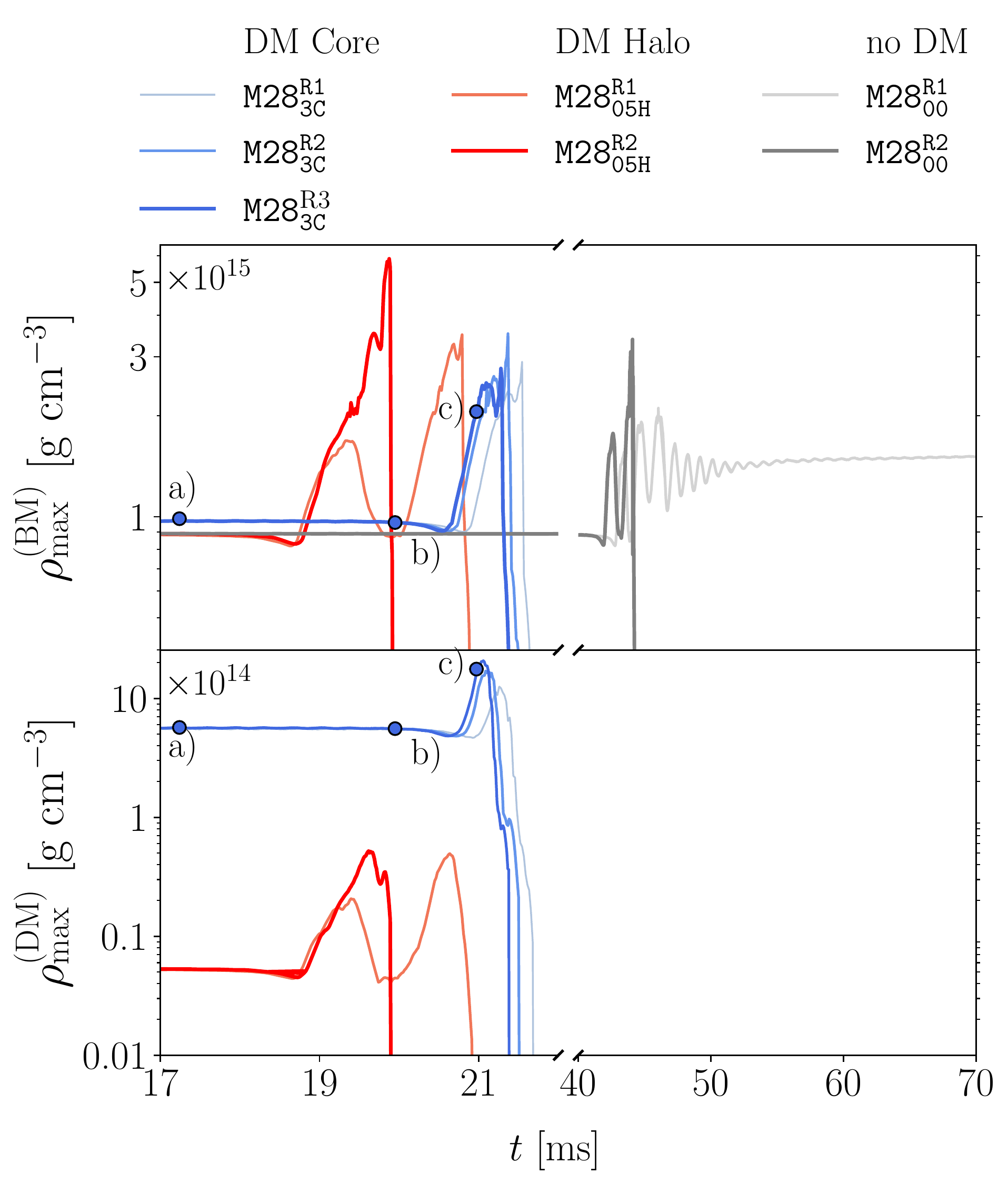}
    \caption{Comparison of the central rest-mass densities during the merger 
    and post-merger phase for BM (upper panel) and DM (lower panel) for the
    $2.8~ M_\odot$ simulations. The central rest-mass density is extracted 
    from the finest refinement level in each run. The points labeled with 
    letters indicate the specific times that correspond to the 2D slices
    shown in Fig.~\ref{fig:panel_core}. Due to a significant decrease in the
    rest-mass density after the collapse to a BH, the central rest-mass 
    density of the fourth panel of Fig.~\ref{fig:panel_core}, \ie{}, 
    point d), lies beyond the plotted range.}
    \label{fig:rho_max_evolution_28}
\end{figure}

The \identifier{M28}{}{} configurations have a different dynamic behavior
characterised by a prompt collapse to a BH.
Figure~\ref{fig:rho_max_evolution_28} shows the evolution of the central rest-mass density of BM and DM for the \identifier{M28}{}{} simulations. During the inspiral, these runs show similar behavior as in the previously discussed setup. Before the merger, we observe minor density oscillations in both fluids. During the merger, DM-halo simulations show a higher peak of the BM density compared to the DM-core or purely baryonic runs, suggesting that a diffuse halo may have an impact on the densities reached during the merger. 
Furthermore, at low resolution ${\tt R1}$, the ${\tt M28}_{\tt 00}^{\tt R1}$
simulation produces an HMNS remnant, while the ${\tt M28}_{\tt 05H}^{\tt R1}$
simulation exhibits a pronounced intermediate peak in the maximal density before reaching
its maximum and and finally collapsing to a BH. This intermediate peak is
indicative of a bounce before merger.
On the other hand, ${\tt M28}_{\tt 3C}^{\tt R1}$ shows a direct increase in
density, collapsing to a BH without any bounce. However, it is worth noting that
the DM-free simulation has a longer inspiral due to the greater initial 
separation between the NSs. This longer inspiral could lead to increased 
numerical diffusion, potentially resulting in mass loss and preventing the
immediate collapse to a BH. At mid-resolution, ${\tt M28}_{\tt 00}^{\tt R2}$
exhibits a BM bounce followed by a collapse to BH. 
The ${\tt M28}_{\tt 05H}^{\tt R2}$ simulation does not
show a bounce, but a gradual increase in the density of BM followed by a 
collapse to a BH. This indicates that the presence of a bounce is not a 
robust feature at this level of resolution.

\begin{figure*}[t]
    \includegraphics[width=0.95\textwidth]{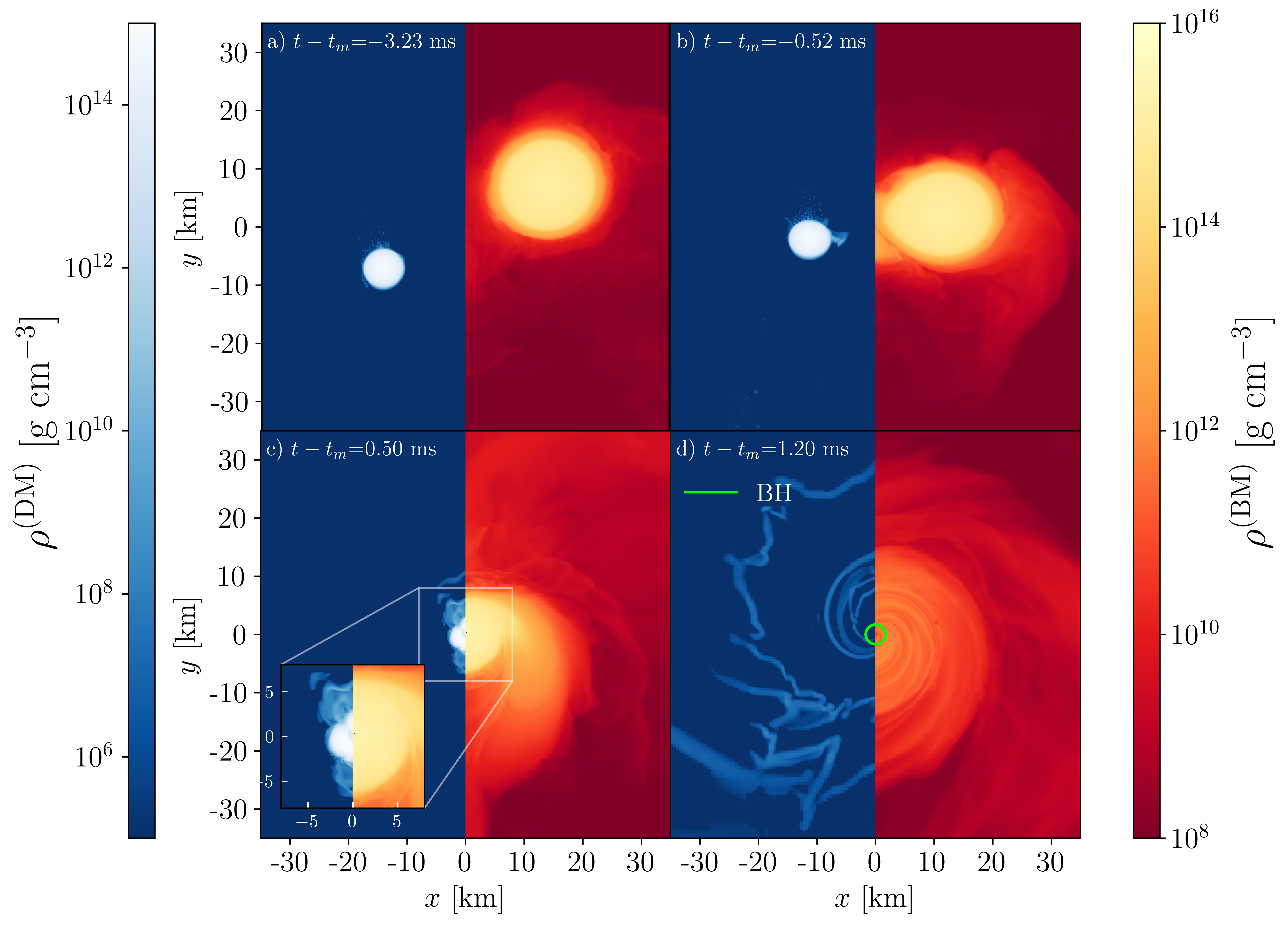}
    \caption{Equatorial density distributions for both BM and DM components, for ${\tt M28}_{\tt 3C}^{\tt R3}$ run at selected times for $M_\mathrm{tot}=2.8 M_\odot$ with $f_\mathrm{DM}=3\%$. The presented times are $t-t_m=(-3.23, -0.52, 0.50, 1.20)$ ms for the upper left, upper right, lower left, and lower right panel, respectively. Those times are the ones marked in Fig.~\ref{fig:rho_max_evolution_28} with the corresponding labels a), b), c) and d). The left halves (negative $x$) show the DM density, whereas the right halves (positive $x$) show the BM
    density.
    }
    \label{fig:panel_core}
\end{figure*}

To further examine the dynamics and better visualize the merger, similar to the
\identifier{M24}{}{} case, we select four key time points a), b), c) and d) that
are marked in Figure~\ref{fig:rho_max_evolution_28} and that correspond to the 
2D slices shown in Figure~\ref{fig:panel_core}. These points represent different
stages of the merger: a) early inspiral, b) formation of a BM/DM bridge linking
the two stars, c) merger time, and d) post-merger phase, showing either the 
newly formed star or a BH.

Figure~\ref{fig:panel_core} shows the spatial distribution of the rest-mass
density for both BM ($\rho^{\rm (BM)}$) and DM ($\rho^{\rm (DM)}$) in the 
\identifier{M28}{R3}{3C}~run, where each baryonic star fully contains a dense DM
core.
At $t-t_m=-3.23$ ms, panel a) shows the two DM-admixed NSs in the early
stage of the inspiral, still relatively widely separated as they continue to
orbit, with an initial central density of 
$\rho_\mathrm{BM, c}=9.729\cdot 10^{14}\, \mathrm{g}/\mathrm{cm}^{3}$ 
and $\rho_\mathrm{DM, c}=5.702\cdot10^{14}\, \mathrm{g}/\mathrm{cm}^{3}$. 
At $t-t_m=-0.52$ ms, as the binary evolves, the stars undergo a significant tidal deformation in both fluids. While the DM remains in the inner regions of the baryonic star, the BM distribution begins to exhibit tidal perturbations due to their mutual gravitational interaction.
A distinct BM bridge forms between the stars at densities around 
$10^{10}\, \mathrm{g}/\mathrm{cm}^{3}$, indicating that the stars came into
contact. On the other hand, the DM cores, even if tidally deformed, remain disconnected structures. This different behavior can be attributed to the significantly larger radius of the BM structure relative to the more compact DM core, resulting in the BM components merging prior to the DM cores. The third snapshot captures the dynamic phase of the BNS merger at $t-t_m=0.50$ ms, panel c) of Figure~\ref{fig:panel_core}. 
The two DM cores have finally merged into a single, central highly dense 
structure with $\rho_\mathrm{max}^{\rm (DM)}=1.735\cdot10^{15}\, \mathrm{g}/\mathrm{cm}^{3}$.
Note, this DM density is still almost an order of magnitude lower than the BM
counterpart, \ie{}, 
$\rho_\mathrm{max}^{\rm (BM)}=7.989\cdot10^{15}\, \mathrm{g}/\mathrm{cm}^{3}$,
indicating that the BM remains the dominant component even in this phase of the merger. The final snapshot ($t-t_m=1.20$ ms, panel d)) shows the post-merger BH, identified by the green circle in the center representing its apparent horizon. Notably, both fluids exhibit a prominent accretion disk structure, whose morphology indicates a significant angular momentum in the remnant. 

\subsection{Remnant properties}\label{subsection:remnantproperties}
Table~\ref{table:remnantprops} presents a summary of the merger outcomes for all simulated configurations. 

We define a prompt collapse as the formation of a BH within the first $2$ ms from the merger time.  For the \identifier{M28}{}{} configurations, a prompt collapse to a BH was observed, except for ${\tt M28}_{\tt 00}^{\tt R1}$, which ends up as an HMNS.
We attribute this discrepancy to the low resolution of the run. 
We report the masses of the formed BHs as the irreducible masses at
$t-t_{m}=25~{\rm ms}$.
For ${\tt M28}_{\tt 00}^{\tt R2}$, the purely baryonic run, the BH mass 
is $2.484 M_\odot$. While in the case of a DM core configuration, 
\ie{}, ${\tt M28}_{\tt 3C}$, the BH masses are $2.519 M_\odot$ 
for \texttt{R1}, $2.537 M_\odot$ for \texttt{R2} and $2.539 M_\odot$ 
for \texttt{R3}, for DM halos, \ie{}, ${\tt M28}_{\tt 05H}$, 
we obtained BH mass of $2.403 M_\odot$ for \texttt{R1} and $2.431 M_\odot$ 
for \texttt{R2}.
Notably, we observe that the BH mass tends to be greater when DM 
is concentrated in a dense core, compared to both DM-free runs 
and configurations with diluted halos.  These results indicate a potential
correlation between the DM morphology and the resulting BH mass. 
We note that the BH in DM-free simulations may have a lower mass due to mass
diffusion related to the longer inspiral. Regardless of this, the
higher BH mass in DM-core configurations can be explained by the fact that
the NSs are more compact and more tightly bound and therefore eject less
mass during the merger.
Moreover, the simulations indicate a possible effect of the DM morphology on the
BH spin.  ${\tt M28}_{\tt 3C}^{\tt R2}$ simulation shows a value of 
$\chi=0.537$.  For the DM-free run ${\tt M28}_{\tt 00}^{\tt R2}$, 
the resulting BH spin is $\chi=0.557$.  Simulations with DM halos, 
\ie{}, ${\tt M28}_{\tt 05H}^{\tt R2}$, yield a spin value of $\chi=0.541$. 
These findings support the potential correlation between the DM configurations and the resulting BH mass and spin.

On the other hand, in $2.4~ M_\odot$ configurations, HMNSs are consistently
formed for all resolutions. In such remnants, a marginal increase in the 
maximum BM density of the HMNS was observed in simulations with a DM core, 
\ie{}, ${\tt M24}_{\tt 3C}$, compared to pure BM simulations and those with DM
halos.  These HMNSs exhibit an enhanced maximum density in the inner regions,
resulting in denser post-merger environments. 
Specifically, at $t-t_m=25$ ms, the maximum BM densities are
$1.179\cdot10^{15}$, $1.152\cdot 10^{15}$ and 
$1.151\cdot10^{15}\, \mathrm{g}/\mathrm{cm}^{3}$ for 
${\tt M24}_{\tt 3C}^{\tt R2}$, ${\tt M24}_{\tt 00}^{\tt R2}$ and 
${\tt M24}_{\tt 05H}^{\tt R2}$, respectively. 
As anticipated, the DM core configurations show higher DM densities, reflecting
their initial denser configuration.

We now turn to the overall post-merger matter distribution. DM-halo simulations
${\tt M24}_{\tt 05H}^{\tt R2}$ show that in the post-merger phase the DM is
distributed throughout the computational domain, maintaining a smooth halo-like
structure. The same behavior of preserving the initial DM morphology is also present for DM cores. In fact, when initially considered concentrated in a core, DM tends to remain spatially constrained in the central region of the remnant. Notably, at $t\lesssim3$ ms after the merger, the DM density distribution forms a ring around the rotational center of the remnant.
Specifically, the peak of the DM distribution is located approximately 
$2.49$ km from the assumed rotational center with 
$\rho_\mathrm{max}^{\rm (DM)}=4.912\cdot 10^{14}\, \mathrm{g}/\mathrm{cm}^{3}$ and $\rho_\mathrm{max}^{\rm (DM)}=4.880\cdot 10^{14}\, \mathrm{g}/\mathrm{cm}^{3}$ for \identifier{M24}{3C}{R2} and \identifier{M24}{3C}{R3}, respectively. 
Here, the center of rotation is considered to be the minimum of the lapse function in the computational domain and will be further discussed in Sec.~\ref{subsection:Angular-velocity}. Later on in the merger, the ${\tt M24}_{\tt 3C}^{\tt R3}$ remnant gradually undergoes a relaxation, resulting in a DM morphology closer to the ID. DM remains concentrated in the central region of the remnant, forming a core with a radius of $\sim6.3$ km, surrounded by a region of low DM density.
\begin{table*}
    \begin{tabular}{lcccccccccc}
    \toprule
        identifier & Remnant & $M_\mathrm{BH}~[M_\odot]$& $\chi_\mathrm{BH}$ &
         $M^{\mathrm{(BM)}}_\mathrm{ejecta}~[M_\odot]$&
         $M^{\mathrm{(DM)}}_\mathrm{ejecta}~[M_\odot]$& 
         $\rho^\mathrm{(BM)}_\mathrm{max}$ [$\mathrm{g}/\mathrm{cm}^{3}$]& 
         $\rho^\mathrm{(DM)}_\mathrm{max}$ [$\mathrm{g}/\mathrm{cm}^{3}$] & $t_m$ [ms] & ~ & ~ \\
        \midrule
        \identifier{M24}{00}{R1}* & HMNS & - & - & 1.76 $\cdot 10^{-2}$ & - & 9.95$\cdot 10^{14}$ & - & 57.09 & ~ & ~ \\ 
        \identifier{M24}{00}{R2}* & HMNS & - & - & 4.27 $\cdot 10^{-3}$ & - & 1.15$\cdot 10^{15}$ & - & 53.65 & ~ & ~ \\ 
        \identifier{M24}{3C}{R1}  & HMNS & - & - & 1.24 $\cdot 10^{-2}$& 2.12$\cdot 10^{-5}$ & 1.14$\cdot 10^{15}$ & 7.20$\cdot 10^{14}$ & 32.47 & ~ & ~ \\
        \identifier{M24}{3C}{R2}  & HMNS & - & - & 6.64 $\cdot 10^{-3}$ & 5.01$\cdot 10^{-6}$ & 1.18$\cdot 10^{15}$ & 7.88$\cdot 10^{14}$ & 30.92 & ~ & ~ \\
        \identifier{M24}{3C}{R3}  & HMNS & - & - & 5.13 $\cdot 10^{-3}$ & 1.41$\cdot 10^{-6}$ & 1.25$\cdot 10^{15}$ & 7.89$\cdot 10^{14}$ & 30.94 & ~ & ~ \\ 
        \identifier{M24}{05H}{R1~} & HMNS & - & - & 3.33 $\cdot 10^{-3}$& 5.70$\cdot 10^{-5}$ & 1.13$\cdot 10^{15}$ & 6.59$\cdot 10^{12}$ & 30.42 & ~ & ~ \\ 
        \identifier{M24}{05H}{R2~} & HMNS & - & - & 2.99 $\cdot 10^{-3}$ & 4.95$\cdot 10^{-5}$ & 1.15$\cdot 10^{15}$ & 7.20$\cdot 10^{12}$ & 30.78 & ~ & ~ \\ 
        
        \identifier{M28}{00}{R1}* & HMNS & - & - & 2.28 $\cdot 10^{-2}$ & - & 1.54$\cdot 10^{15}$ & - & 44.29 & ~ & ~ \\ 
        \identifier{M28}{00}{R2}* & BH & 2.484 & 0.557 & 4.62 $\cdot 10^{-3}$& - & 1.28$\cdot 10^{11}$ & - & 43.43 & ~ & ~ \\ 
        \identifier{M28}{3C}{R1} & BH & 2.519 & 0.530 & 1.52 $\cdot 10^{-3}$ & 0 & 3.53$\cdot 10^{10}$ & 5.68$\cdot 10^{3}$ & 22.32 & ~ & ~ \\ 
        \identifier{M28}{3C}{R2} & BH & 2.537 & 0.537 & 2.06 $\cdot 10^{-4}$ & 0 & 1.10$\cdot 10^{10}$ & 5.70$\cdot 10^{3}$ & 22.15 & ~ & ~ \\
        \identifier{M28}{3C}{R3} & BH & 2.539 & 0.538 & 1.76$\cdot 10^{-5}$ & 0 & 2.65$\cdot 10^{9}$ & 5.71$\cdot 10^{3}$ & 22.08 & ~ & ~ \\ 
        \identifier{M28}{05H}{R1~} & BH & 2.403 & 0.495 & 3.89$\cdot 10^{-3}$ & 3.57$\cdot 10^{-4}$ & 2.86$\cdot 10^{9}$ & 1.04$\cdot 10^{10}$ & 20.15 & ~ & ~ \\
        \identifier{M28}{05H}{R2~} & BH & 2.431 & 0.541 & 1.59 $\cdot 10^{-3}$ & 2.51$\cdot 10^{-6}$ & 2.38$\cdot 10^{8}$ & 2.38$\cdot 10^{8}$& 20.23 & ~ & ~ \\ 
        \bottomrule
    \end{tabular}\caption{Overview of post-merger quantities at $t-t_m=25$ ms. The table lists simulation identifiers, remnant fate, 
    irreducible mass $M_\mathrm{BH}$ and spin $\chi_\mathrm{BH}$, BM and DM ejecta masses
    $M^{\mathrm{(BM)}}_\mathrm{ejecta}$ and $M^{\mathrm{(DM)}}_\mathrm{ejecta}$
    (extracted at $295~{\rm km}$ on the finest grid level present at such distance),
    maximum BM and DM densities $\rho^\mathrm{(BM)}_\mathrm{max}$ and
    $\rho^\mathrm{(DM)}_\mathrm{max}$, evaluated from the finest grid level 
    of the simulation, and merger time $t_m$.
    Starred identifiers, \egabb{} \identifier{M24}{00}{R1}*, 
    have a larger initial separation distance, resulting in an offset 
    for the merger times.}\label{table:remnantprops}
\end{table*}

\subsection{Angular-velocity evolution}\label{subsection:Angular-velocity}
To better understand the dynamic interplay between the two fluids, we analyze
the angular velocity profiles of the HMNS remnants. We define the
angular velocity $\Omega^{(s)}$ as
\begin{equation}\label{eq:omega_defGR}
    \Omega^{(s)} \equiv \frac{d\phi}{dt} = \frac{dx^\phi}{dt} = \frac{u^{(s)\phi}}{u^{(s) t}}\,,
\end{equation}
where $u^{(s) t}$ and $u^{(s)\phi}$ are the temporal and angular components 
of the four-velocity vector $u^{(s)\mu}$. The corresponding three-velocity, 
as measured by the Eulerian observer, can be evaluated as
\begin{equation}
    v^{(s)i} \equiv \frac{\gamma^i_\mu u^{(s)\mu}}{-n_\mu u ^{(s)\mu}}=\frac{1}{\alpha}\left(\frac{u^{(s)i}}{u^{(s)t}}+\beta^i \right)\,,
\end{equation}
where $n_\mu$ is the timelike normal defined in Eq.~\eqref{eq:timelike_normal}.
The angular velocity from Eq.~\eqref{eq:omega_defGR} within the 3+1 formalism is thus given by
\begin{equation}\label{eq:omega_def}
    \Omega^{(s)} = \alpha v^{(s)\phi} -\beta^\phi\,,
\end{equation}
where the angular components are obtained through
\begin{align}
    v^{(s)\phi} ={}& \frac{(x-x_0)v^{(s)y} -(y-y_0)v^{(s)x} }{(x-x_0)^2+(y-y_0)^2}\,,\\
     \beta^\phi ={}& \frac{(x-x_0)\beta^y-(y-y_0)\beta^x}{(x-x_0)^2+(y-y_0)^2}\,,
\end{align}
are the three-velocity and the shift vector components computed from the 
Cartesian grid with coordinates $(x, y, z)$ for a given center of rotation 
$(x_0, y_0, z)$. Eq.~\eqref{eq:omega_def} can be seen as the lapse-corrected 
part of the $\phi$-component of the three-velocity minus a frame-dragging term
provided by the $\phi$-component of the shift vector~\cite{Hanauske:2016gia}. 
We then define the azimuthally-averaged angular velocity $\bar{\Omega}^{(s)}$ as
\begin{equation}
    \bar{\Omega}^{(s)}(r, t)=\int_{-\pi}^{\pi}\Omega^{(s)} (r, \phi, t)d\phi\,.
\end{equation}

A significant aspect of this analysis is the uncertainty in determining the
precise center of rotation for each fluid component within the simulations. 
Due to the symmetry of our setup the center of rotation should be located
at the origin, however, due to accumulated numerical round-off error the actual
center of rotation is slightly displaced from the origin.
Furthermore, due to the distinct distributions of the BM and DM components, 
the precise center of rotation is inherently uncertain and might not coincide
for both fluid components. To address this, the minimum of the lapse function $\alpha(x,y,z)$ within the simulation is used as the coordinate center, recognizing that it might not perfectly represent the rotational center of each component. To account for potential variations in the true rotational center relative to this coordinate center, we introduce a parameter 
$\Delta^{(s)}$, which represents the distance between the chosen center 
and the location of the lapse minimum.
We generate a uniform sample of 500 centers located in the equatorial plane
distributed within a disk with a radius of 0.5~km from the location of the 
lapse minimum. For each of these 500 points, we then generate the
associated $\bar{\Omega}^{(s)}$ profiles. Given that these DM-admixed NS are
predominantly composed of BM, the minimum of the lapse function is expected to
track the BM rotational center more closely, leading to lower expected uncertainty in the BM profiles compared to the DM ones. 

\begin{figure*}
    \includegraphics[width=0.95\textwidth]{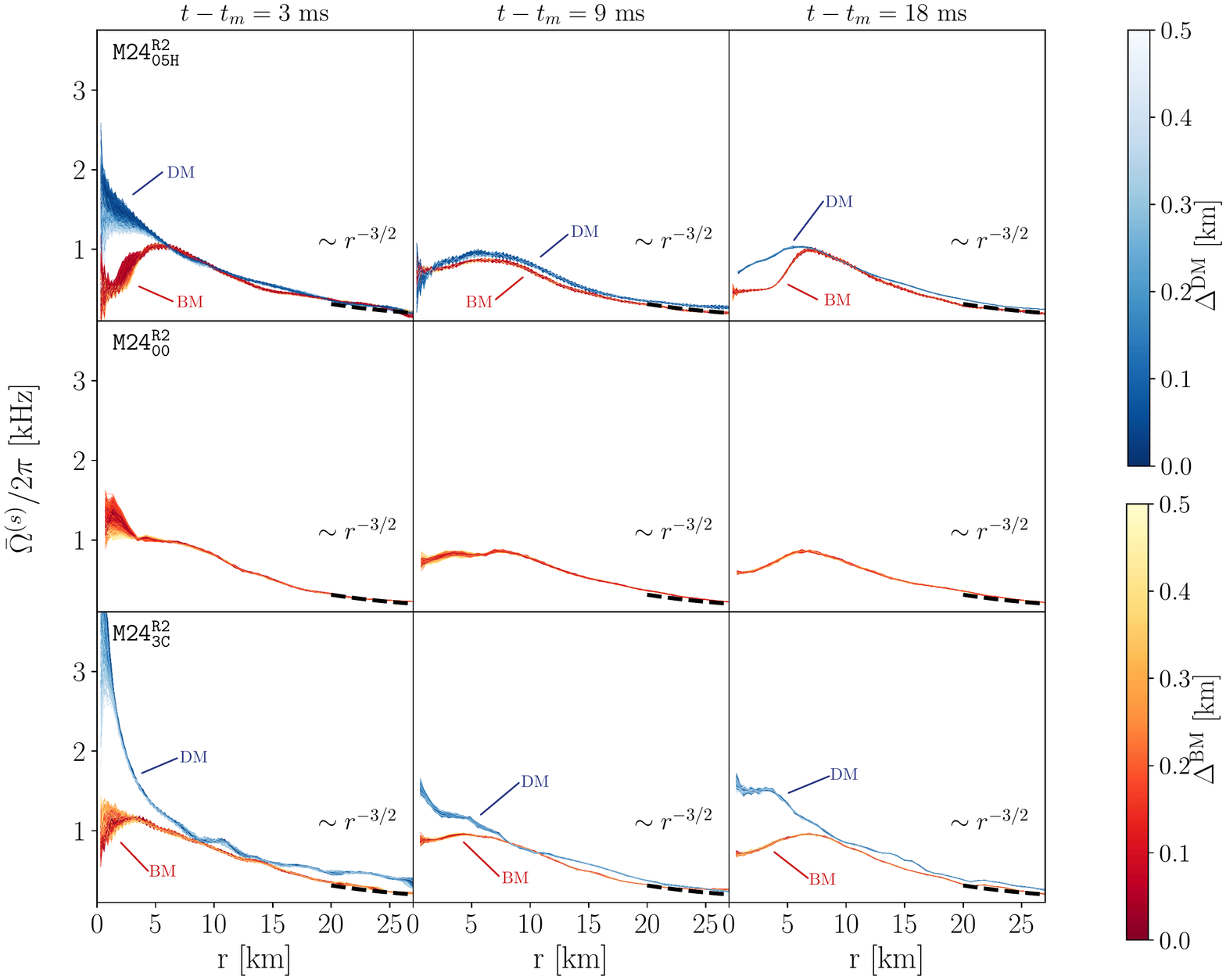}
    \caption{Azimuthally-averaged angular velocity $\bar{\Omega}^{(s)}(r, t)$ profiles on the equatorial plane for the ${\tt M24}^{\tt R2}$ simulations on the finest grid level. The parameter $\Delta^{(s)}$ represents the distance between a chosen rotational center and the lapse-informed coordinate center. Dashed black lines serve as a reference profile scaling like $\sim r^{-3/2}$.
  }\label{fig:omegar}
\end{figure*}

Figure~\ref{fig:omegar} presents the radial profiles of the averaged fluid angular velocity, $\bar{\Omega}^{(s)}(r, t)$, on the equatorial plane for ${\tt M24}^{\tt R2}$ runs, those with an HMNS remnant, without a BH collapse observed within the simulation time, at three different times after the merger $t-t_m=3$, 6 and 9~ms.

At $t-t_m=3$ ms, we observe quantifiable differences in the angular velocity profiles. At such early times after the merger, as the angular momentum is redistributed outward, the angular velocity profiles of both fluids exhibit
significant uncertainties, mainly in the closest-to-the-center regions. 
Over time the angular momentum redistribution tends to reduce the central angular
velocity of each fluid, creating different rotational zones within the stellar
structure. 
In simulations with a diffuse DM halo, \ie{} ${\tt M24}_{\tt 05H}^{\tt R2}$, 
the DM component shows an angular velocity gradient, with angular velocity
maxima distributed in the interval $[0.92, 2.59]$~kHz at the considered rotational center, depending on the 
considered spatial offset $\Delta^{(\mathrm{DM})}$.
On the other hand, the DM core configuration shows a rapidly rotating DM 
sub-structure immediately after the merger. The DM angular velocities
significantly exceed their BM counterpart, having their maximum angular 
velocities distributed in the interval $[1.95, 4.73]$~kHz at the considered rotational center.
Close to the center the BM angular velocity profiles of all configurations
differ. The smallest central angular velocities is observed for 
\identifier{M24}{R2~}{05H}.
To provide a representative measure of the central angular velocity
we average over all sampled rotation centers, 
yielding for \identifier{M24}{R2~}{05H} an average central angular velocity of
$\bar{\Omega}^{\mathrm{(BM)}}\simeq0.5$~kHz,
whereas for \identifier{M24}{R2}{00} and \identifier{M24}{R2}{3C}
we obtain $\bar{\Omega}^{\mathrm{(BM)}}\simeq1.3$~kHz and $1.0$~KHz,
respectively.
This observation suggests that right after merger the dynamics of the BM
is strongly altered by the relatively small amount of DM.

As time evolves, at $t-t_m=9$ ms, the central DM angular velocity exhibits significant damping due to redistribution of the angular momentum. In DM-admixed HMNSs, we observed a substantial reduction in the central DM angular velocity $\bar{\Omega}^\mathrm{(DM)}$. Notably, in the ${\tt M24}_{\tt 05H}^{\tt R2}$
simulation, both DM and BM match closely. However, the DM component maintains a
slightly higher rotational frequency, with peak values of $0.873$~kHz and 
$0.966$~kHz for DM and BM, respectively. 
However, in the DM core configuration, the dense DM core still rotates 
considerably faster than the BM fluid, showing a higher angular velocity even 
when compared to the DM halo configuration, with profiles even peaking at
different radii, as shown in Table~\ref{tab:omegar}.

At later times, $t-t_m=18$ ms, BM and DM distributions move towards a more 
stable equilibrium. In the ${\tt M24}_{\tt 05H}^{\tt R2}$ simulation, the DM
profile shows slight variations after the previously considered time steps.
However, these profiles reveal a more complex rotational structure within the 
${\tt M24}_{\tt 3C}^{\tt R2}$ simulation. The DM core displays two distinct
rotational regimes. The central region has a near-rigid rotation with an almost
constant frequency. Beyond the DM core boundary, a gradual decrease in 
$\bar\Omega^{(s)}$ is observed at larger radii. Without DM, the BM profiles of 
${\tt M24}_{\tt 00}^{\tt R2}$, the BM profiles appear to approach a stable state.
The remnant shows a deceleration along with a minor inward shift of the angular velocity peak. A qualitative comparison between ${\tt M24}_{\tt 3C}^{\tt R2}$ and ${\tt M24}_{\tt 00}^{\tt R2}$ indicates similarities in their profiles and behavior. However, BM profiles of ${\tt M24}_{\tt 05H}^{\tt R2}$ exhibit a central plateau. In the inner stellar regions, the fluid maintains a relatively uniform angular velocity $\bar\Omega^{(\mathrm{BM})}=0.48$ kHz up to a radius of $\sim3.9$ km, followed by a marked increase that reaches its peak 
$\bar\Omega^{(\mathrm{BM})}=1.002$ kHz at $\sim6.7$ km, remarkably similar to 
the DM one $\bar\Omega^{(\mathrm{BM})}=1.026$ kHz at $\sim5.5$ km. 

All configurations show the expected behavior at asymptotically large radii,
decreasing monotonically as $\propto r^{-3/2}$~\cite{Hanauske:2016gia}, as
indicated by the good agreement with the reference profile (black dashed line)
in Fig.~\ref{fig:omegar}.
\begin{table*}
    \begin{tabular}{lccccc}
    \toprule
    identifier  & $t-t_m$ [ms] & $\bar\Omega^{\mathrm{(BM)}}_\mathrm{max}$ [kHz] & $\bar\Omega^{\mathrm{(DM)}}_\mathrm{max}$ [kHz] & $r_{\bar\Omega^{\mathrm{(BM)}}_\mathrm{max}}$ [km] & $r_{\bar\Omega^{\mathrm{(DM)}}_\mathrm{max}}$ [km] \\
    \midrule
                     & 3    & 1.285   & -       & 1.106  & -       \\
 ${\tt M24}_{\tt00}^{\tt R2}$  & 9    & 0.871   & -       & 6.991  & -       \\
                     & 18   & 0.854   & -       & 6.461  & -       \\ \hline

                                & 3    & 1.062   & 1.736   & 5.376  & 0.436   \\  
${\tt M24}_{\tt05H}^{\tt R2}$  & 9    & 0.873   & 0.966   & 5.257  & 5.876   \\
                                & 18   & 1.002   & 1.026   & 6.762  & 5.460  \\\hline
                     
                     & 3    & 1.153   & 3.961   & 2.238  & 0.350   \\
${\tt M24}_{\tt3C}^{\tt R2}$   & 9    & 0.967   & 1.629   & 2.896  & 0.479   \\
                     & 18   & 0.954  & 1.578   & 6.70   & 0.522   \\
    \bottomrule
    \end{tabular}\caption{Maximum angular velocities $\bar\Omega^{(s)}_{\mathrm{max}}$ and their corresponding radial distances $r_{\bar\Omega^{(s)}_{\mathrm{max}}}$ for both BM and DM at different times $t-t_m$. The $\bar\Omega^{(s)}_{\mathrm{max}}$ values are averaged over all possible values of parameter $\Delta^{(s)}$.}\label{tab:omegar}
\end{table*}
\subsection{Ejecta Masses}\label{subsection:ejecta}
\begin{figure}
\includegraphics[width=\columnwidth]{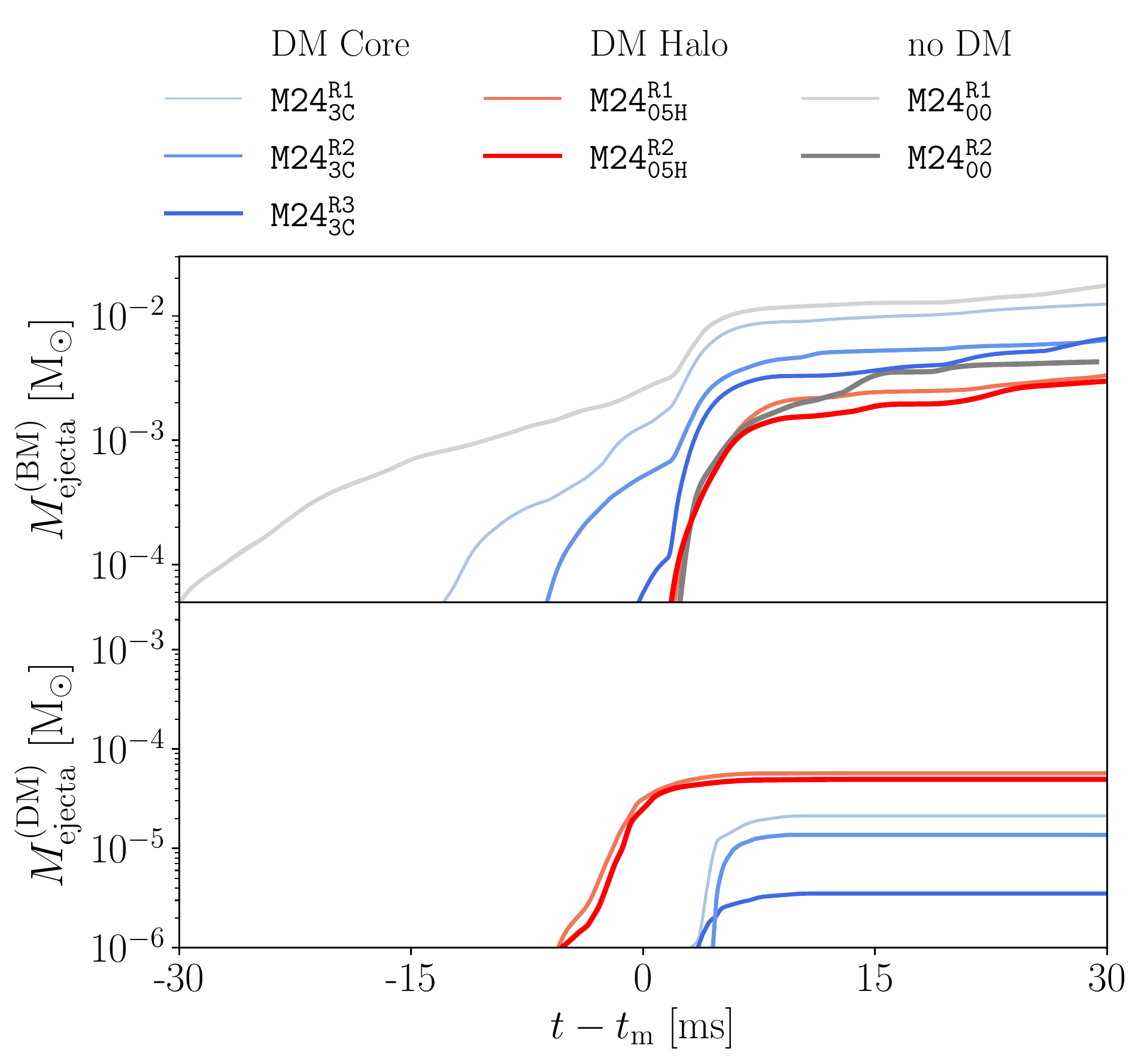}
    \caption{The ejecta mass as a function of time for the 
    \identifier{M24}{}{} runs. The upper panels show the BM ejecta, while the lower panels present the DM ejecta. All ejecta curves shown here are extracted at the smallest extraction radius, \ie{}, $r=295$ km, on the finest level available.
    }
    \label{fig:ejecta_panel}
\end{figure}

\begin{figure}
\includegraphics[width=\columnwidth]{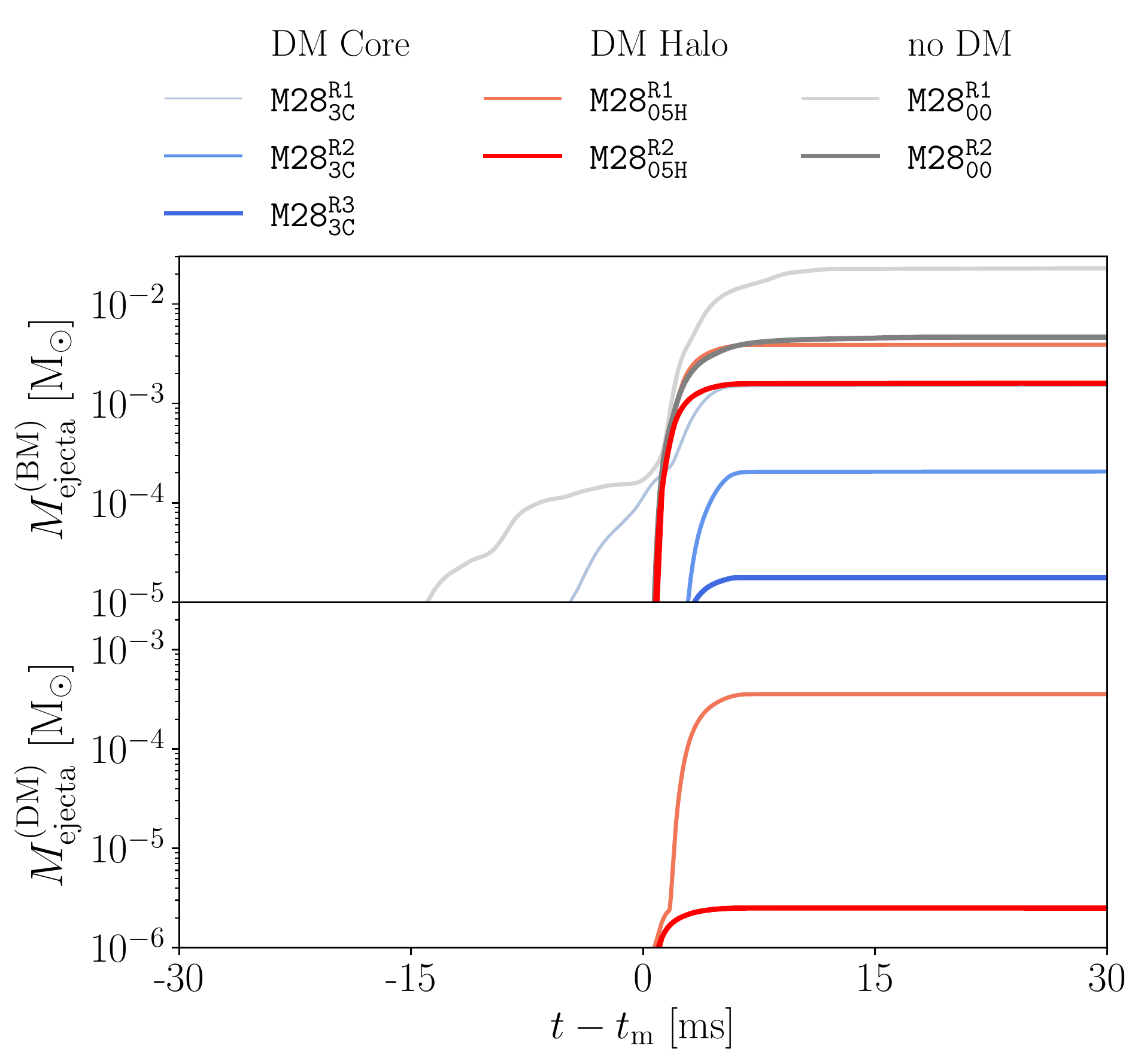}
    \caption{The ejecta mass as a function of time for the \identifier{M28}{}{} 
    runs. The upper panels show the BM ejecta, while the lower panels present 
    the DM ejecta. All ejecta curves shown here are extracted at the smallest
    extraction radius, \ie{}, $r=295$ km, on the finest level available.
    }
    \label{fig:ejecta_panel2}
\end{figure}

In this work, the ejecta properties are determined by computing them on a series
of concentric extraction spheres, with radii ranging between 295~km and 1030~km. On each spherical surface, we computed the integrated fluxes of rest mass,
energy, and momentum for outgoing unbound matter. The matter is considered 
unbound on these extraction surfaces if it satisfies the geodesic
criterion~\cite{Hotokezaka:2012ze}:
\begin{equation}
    u_t <-1\ \mathrm{and}\ v_r > 0\,,
\end{equation} 
where $u_t$ and $v_r$ are, respectively, the time component of the fluid four-velocity and the radial velocity. Figures~\ref{fig:ejecta_panel} and~\ref{fig:ejecta_panel2} show the time evolution of the ejecta mass for \identifier{M24}{}{} and \identifier{M28}{}{}, respectively. 
Table~\ref{table:remnantprops} provides an overview over the ejecta masses
at $t-t_m = 25\,\mathrm{ms}$ extracted at a radius of 295 km.

\subsubsection{BM ejecta}

The values of the BM ejecta masses increase sharply around the merger and then
transition to a gradual increase for all setups. However, as soon as the BH 
forms the ejecta are suppressed and the ejecta mass reaches a stable plateau.

Compared to DM-free setups DM halos lead to a suppression of BM ejecta
by a factor of approximately 10.
This suppression of ejecta can be intuitively understood by the deeper 
potential well of the DM-enhanced gravitational field.

For DM cores on the other hand there are large differences between configurations.
We find evidence that in higher mass systems, a dense DM core can actively
suppress the ejection of BM, while in lower mass system the core causes an 
increase in BM ejecta, probably due to a more violent merger and post-merger
dynamics. 
The higher mass system of \identifier{M28}{3C}{~~} exhibits a supression of BM
ejecta by a factor 100, while for the lower mass system 
\identifier{M24}{3C}{~~} DM ejecta are actually slightly larger than in the
\identifier{M24}{00}{R2} configuration.
This observation is in accordance with the previous study of~\citet{Emma:2022xjs}, 
where the
authors observed that, for a configuration with the same mass and DM fraction
as \identifier{M24}{3C}{~~}, the BM ejection was enhanced.

While the comparisons between resolutions show that in some cases the amount of
ejecta is sensitive to the numerics, the consistent trend across 
resolutions indicates that the overall picture of the DM's impact on the ejecta
dynamics is robust and not significantly altered by numerical artifacts.

\subsubsection{DM ejecta}
Beyond the BM ejecta, our simulations also reveal the presence of DM ejecta,
providing direct insight into its dynamic behavior during these mergers. 
In the \identifier{M24}{}{} configurations, for DM halos, the ejecta mass of
\identifier{M24}{05H}{R2} reaches a plateau around $5\cdot10^{-5} M_\odot$. 
In comparison, DM cores show reduced ejecta masses, saturating at 
$5\cdot10^{-6} M_\odot$ for \identifier{M24}{3C}{R2}. This suggests that diluted
configurations lead to significantly greater DM ejecta compared to the dense DM
core scenario.

In the \identifier{M28}{}{} configurations, the DM ejecta mass shows a similar trend. While the DM halos show a rapid increase of ejecta mass, reaching a plateau around $3\cdot10^{-6} M_\odot$ for \identifier{M28}{05H}{R2}, the DM core simulations show a complete absence of DM ejecta. This behavior demonstrates that prompt collapse in such massive systems, when coupled with a dense DM core, results in the complete capture of DM by the forming BH, effectively halting the ejection.

\subsection{Gravitational Waves}\label{subsection:GW}

\begin{figure*}
  \begin{center}
  \includegraphics[width=0.95\textwidth]{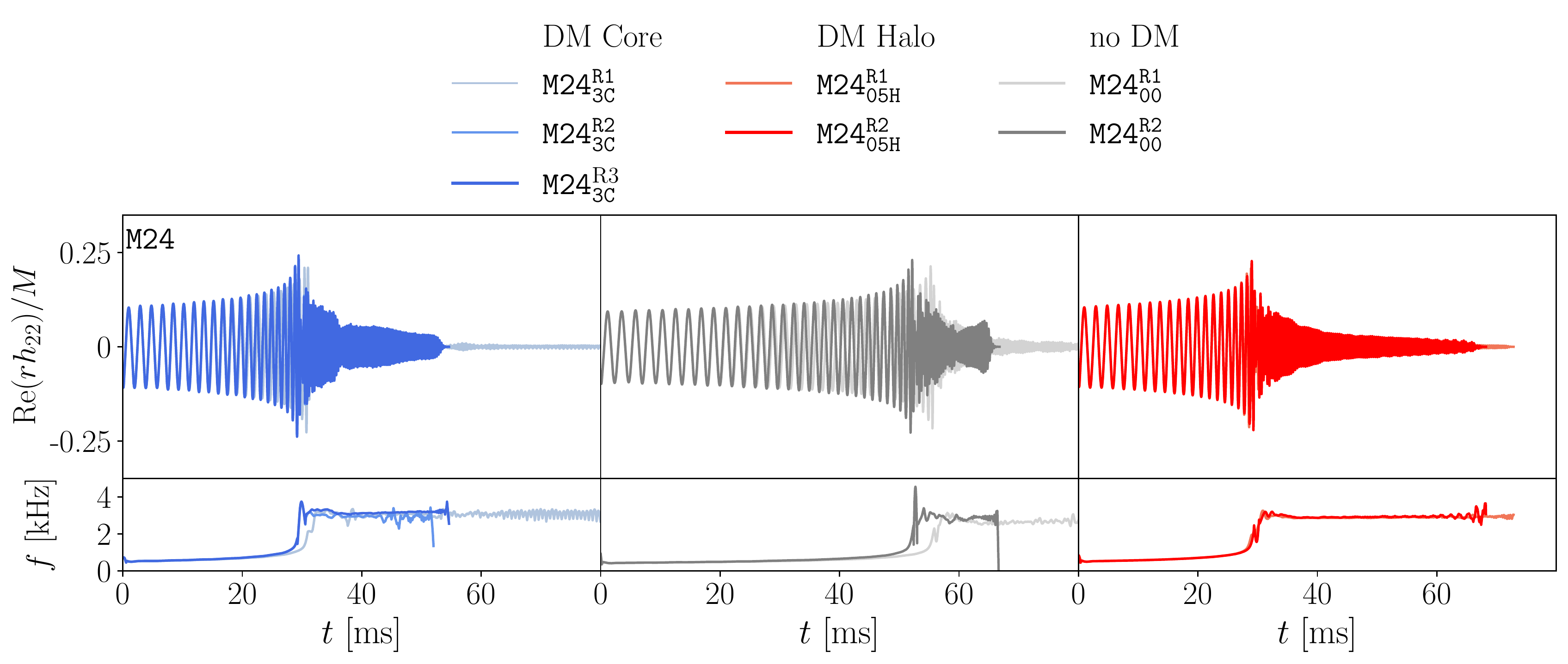} \label{fig:GW_24} \\
  \includegraphics[width=0.95\textwidth]{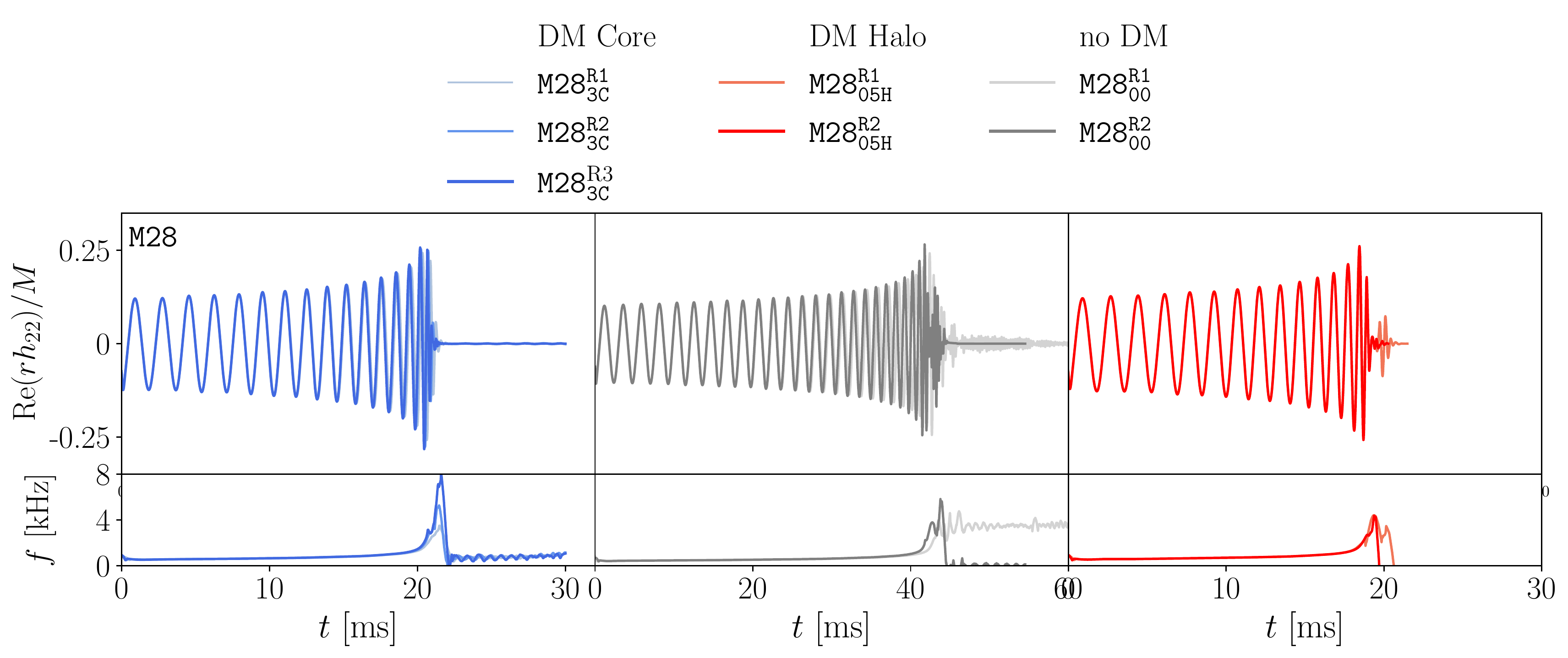} \label{fig:GW_28}\\  
  \end{center}
  \caption{GW waveform strain and instantaneous frequency of the $\ell=|m|=2$ mode
  GW signal for all the simulations run. The instantaneous frequency is computed via Eq.~(53) of Ref.~\cite{Bruegmann:2006ulg}. All the GW waveforms are extracted at $r_\mathrm{ext}=1474.21$~km.
  \textbf{Top panel:} \identifier{M24}{}{} simulations. 
  \textbf{Bottom panel:} \identifier{M28}{}{} simulations. 
  } \label{fig:AB}
\end{figure*}

In this section, we extract the GW signals from the simulations at different extraction radii $r_{\rm extr}$, using the approximate retarded time coordinate~\cite{Thierfelder:2011yi}
\begin{equation}
\label{eq:retarded_time}
		u = t - r_{\rm{ext}} - 2 M_{\rm ADM} \ln \left[\frac{r_{\rm{ext}}}{2M_{\rm ADM}} - 1\right]\,,
\end{equation}
where $t$ is the simulation time. We show the results of the extracted GWs in terms of the dominant $(2,|2|)$-mode strain for different DM configurations and resolutions at a fixed $r_{\rm ext}$ in Fig.~\ref{fig:AB}.

We then compare the GW signals (extracted at a common largest $r_{\rm ext}$ per configuration) resulting from the NR simulations with an existing GW model for BNS systems without DM. In this case, we use the \imrphenomxasnrtidalthree{} model~\cite{Abac:2023ujg, lalsuite} for the comparisons. This model employs the \texttt{IMRPhenomXAS} model for the dominant $(2,|2|)$ mode GWs from aligned-spin binary black hole (BBH) systems~\cite{Pratten:2020fqn}. On the other hand, \texttt{NRTidalv3} is the latest version of the modular, phenomenological NRTidal series~\cite{Dietrich:2017aum, Dietrich:2018uni, Dietrich:2019kaq} that describes the tidal effects that are prevalent for BNS systems, which arise from the tidal interactions between the NSs. 

For the comparison, we align the NR waveforms of different configurations 
with the corresponding \imrphenomxasnrtidalthree{} waveform by computing 
the time and phase shifts $\delta t$ and $\delta \phi$ that minimize the 
following integral~\cite{Hotokezaka:2015xka}:%
\begin{equation}
\mathcal{I}(\delta t, \delta \phi) = \int_{t_1}^{t_2} dt |\phi_{\text{NR}}(t) - \phi_{\text{Model}}(t + \delta t) + \delta \phi|\,, 
\end{equation}
over some alignment window bounded by the time interval $[t_1, t_2]$, which is chosen near the beginning of the NR waveform.

\begin{figure*}
\includegraphics[width=0.47\linewidth]{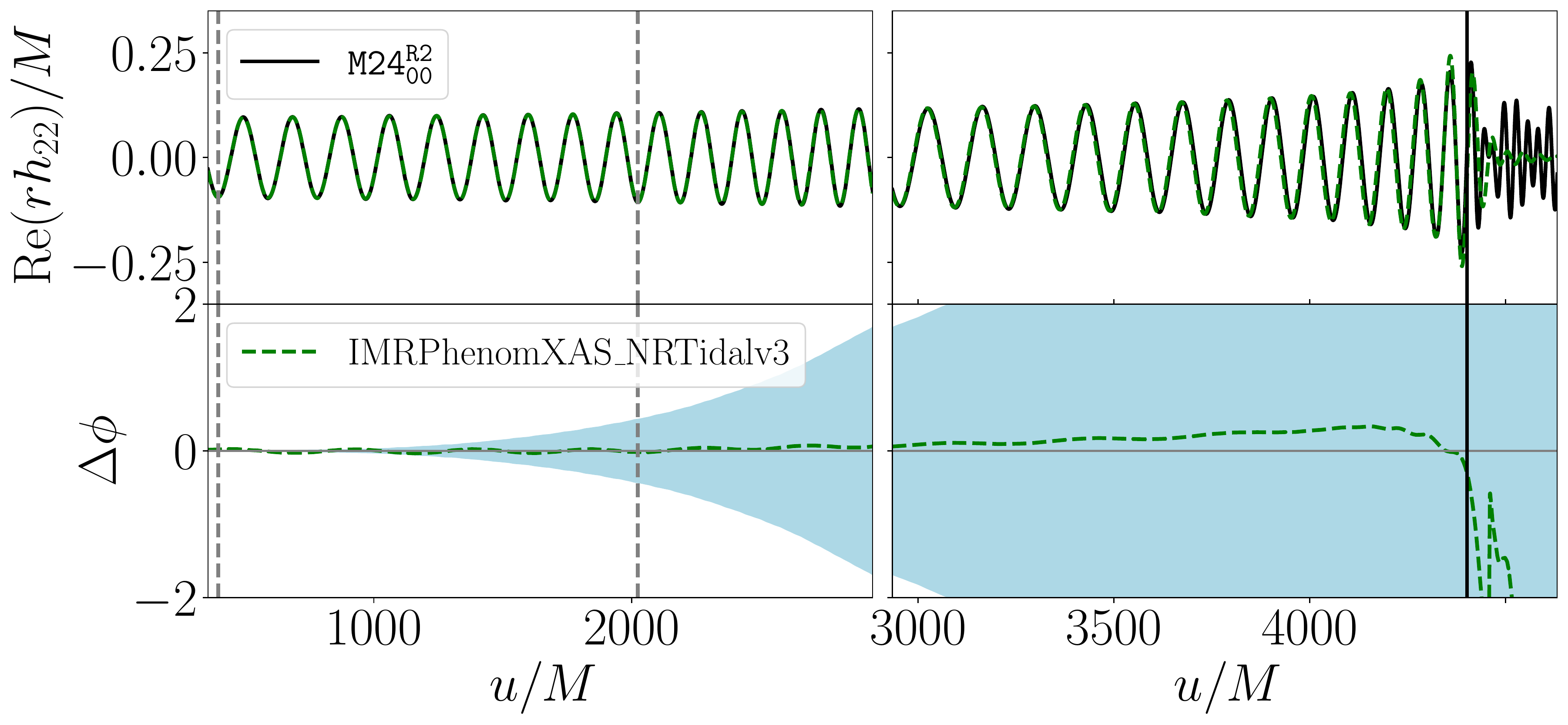}\hfill
\includegraphics[width=0.47\linewidth]{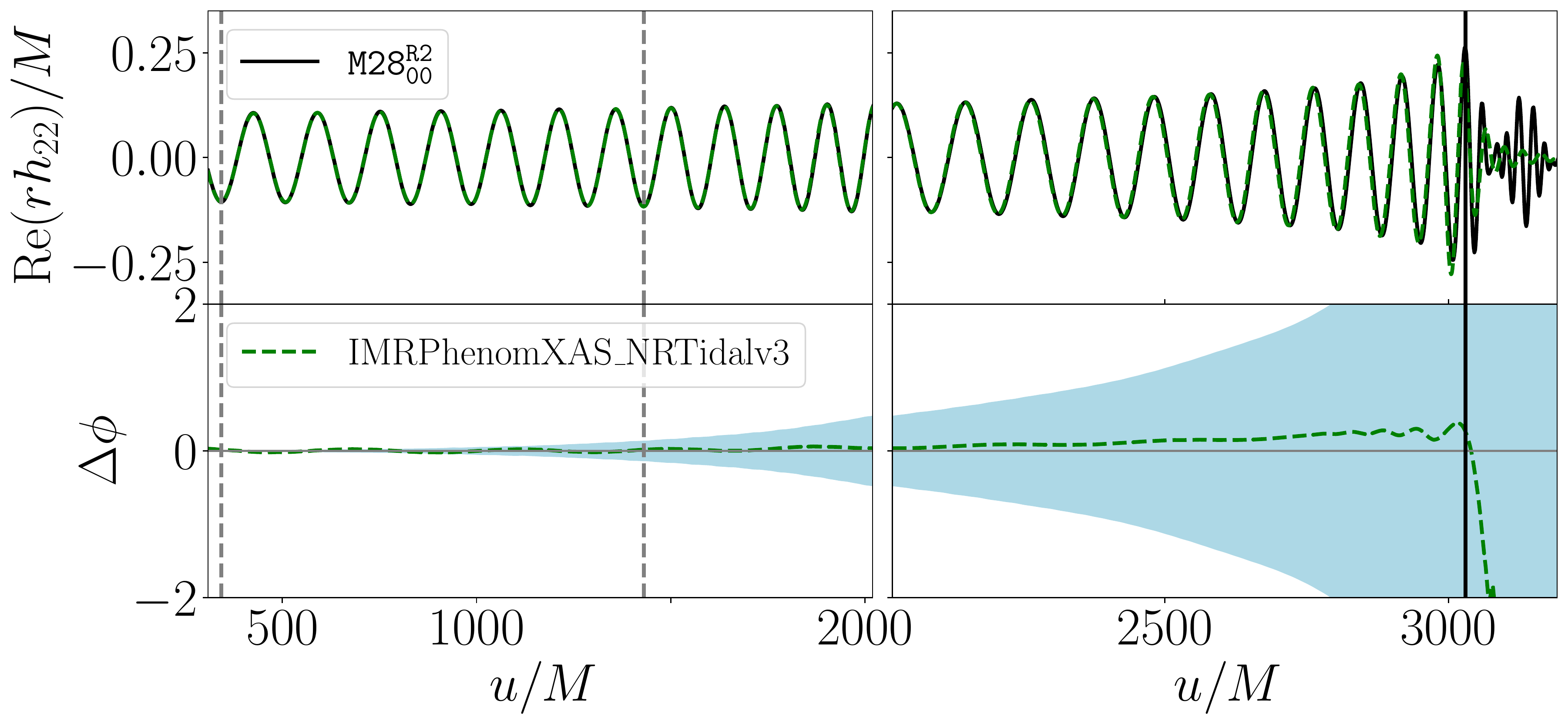}
\includegraphics[width=0.47\linewidth]{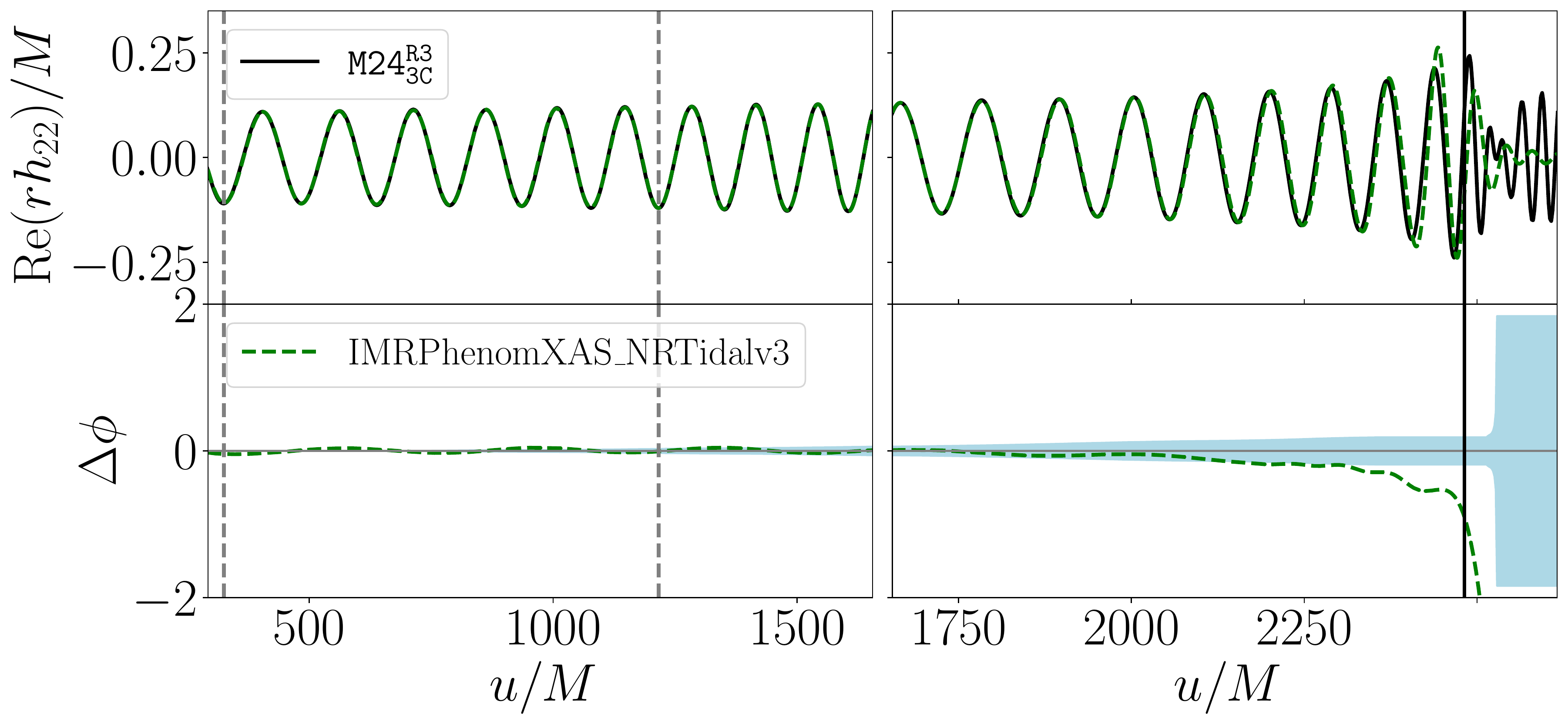}\hfill
\includegraphics[width=0.47\linewidth]{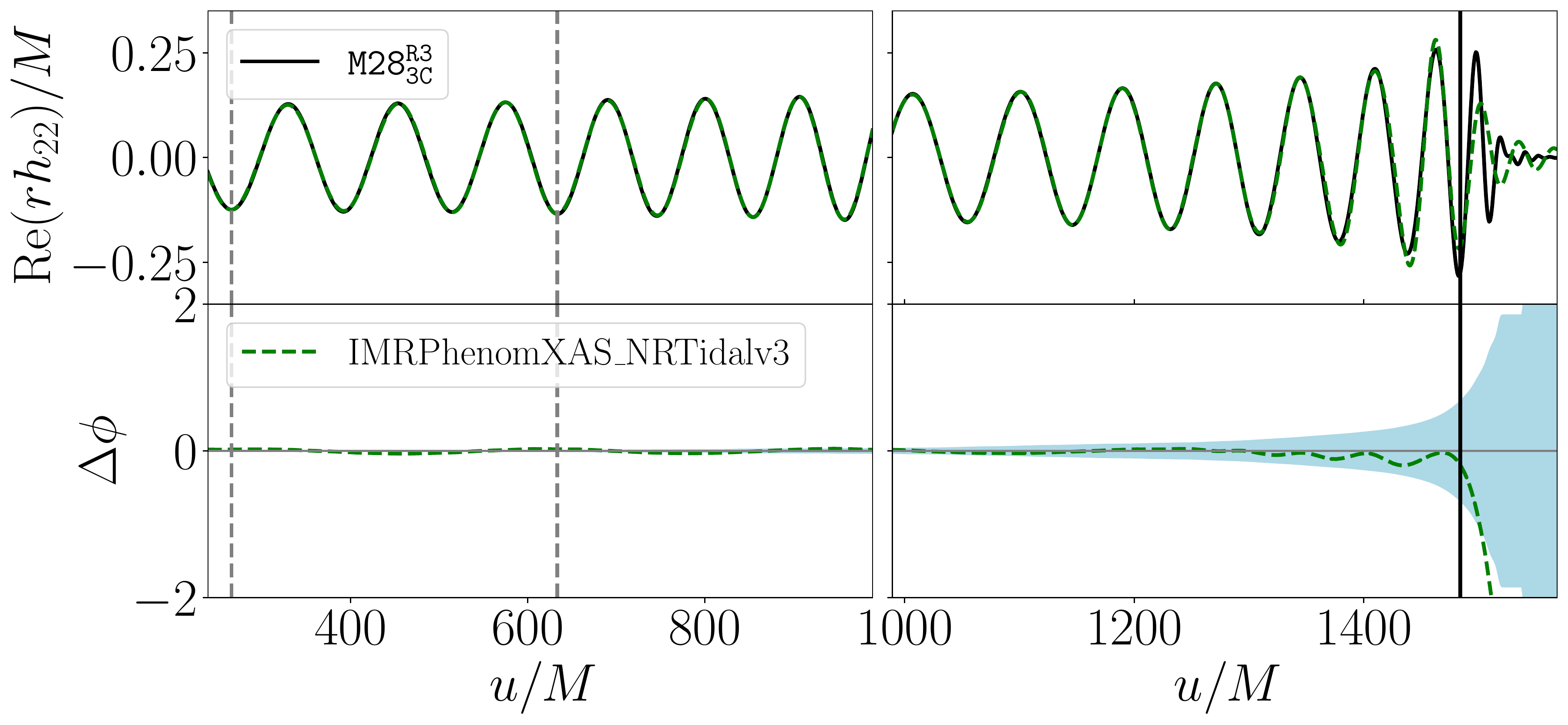}
\includegraphics[width=0.47\linewidth]{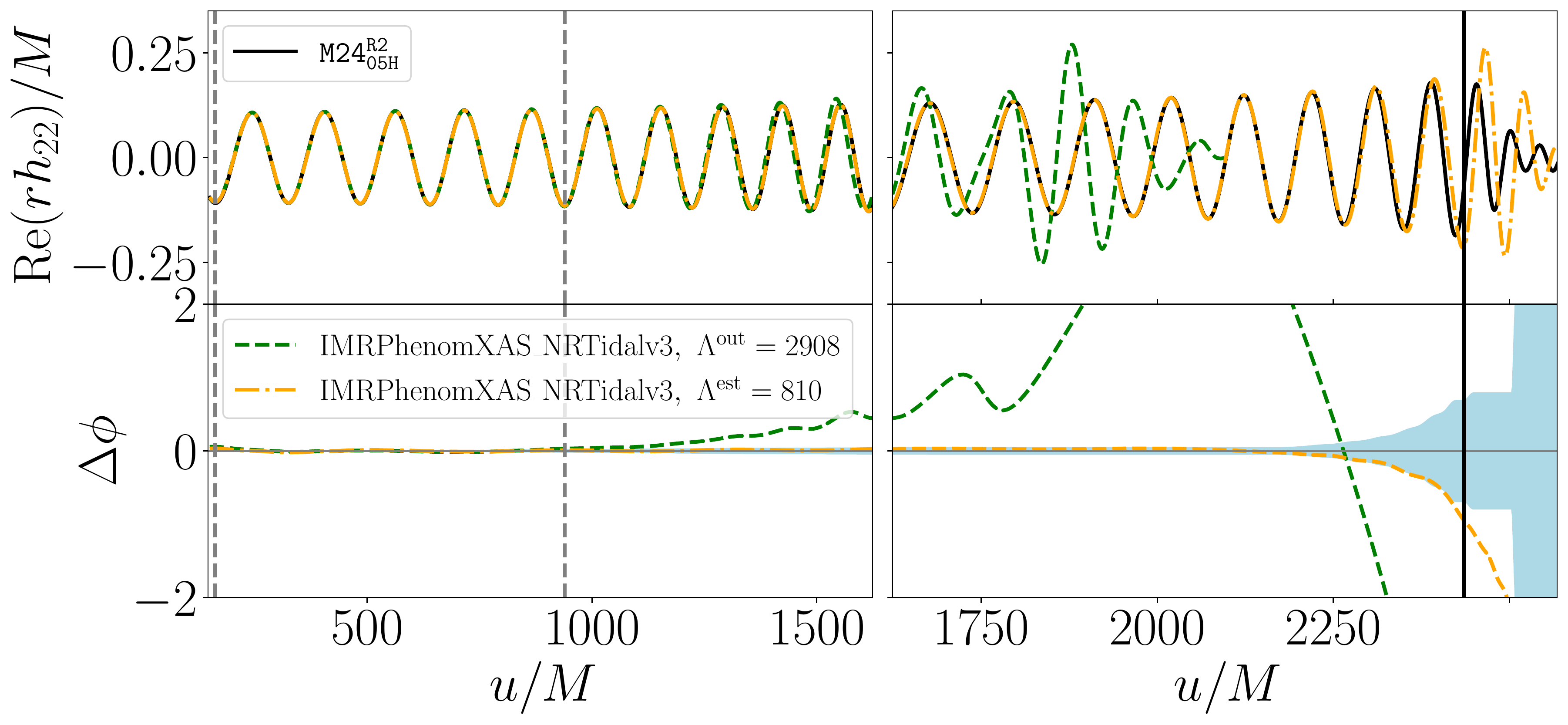}\hfill
\includegraphics[width=0.47\linewidth]{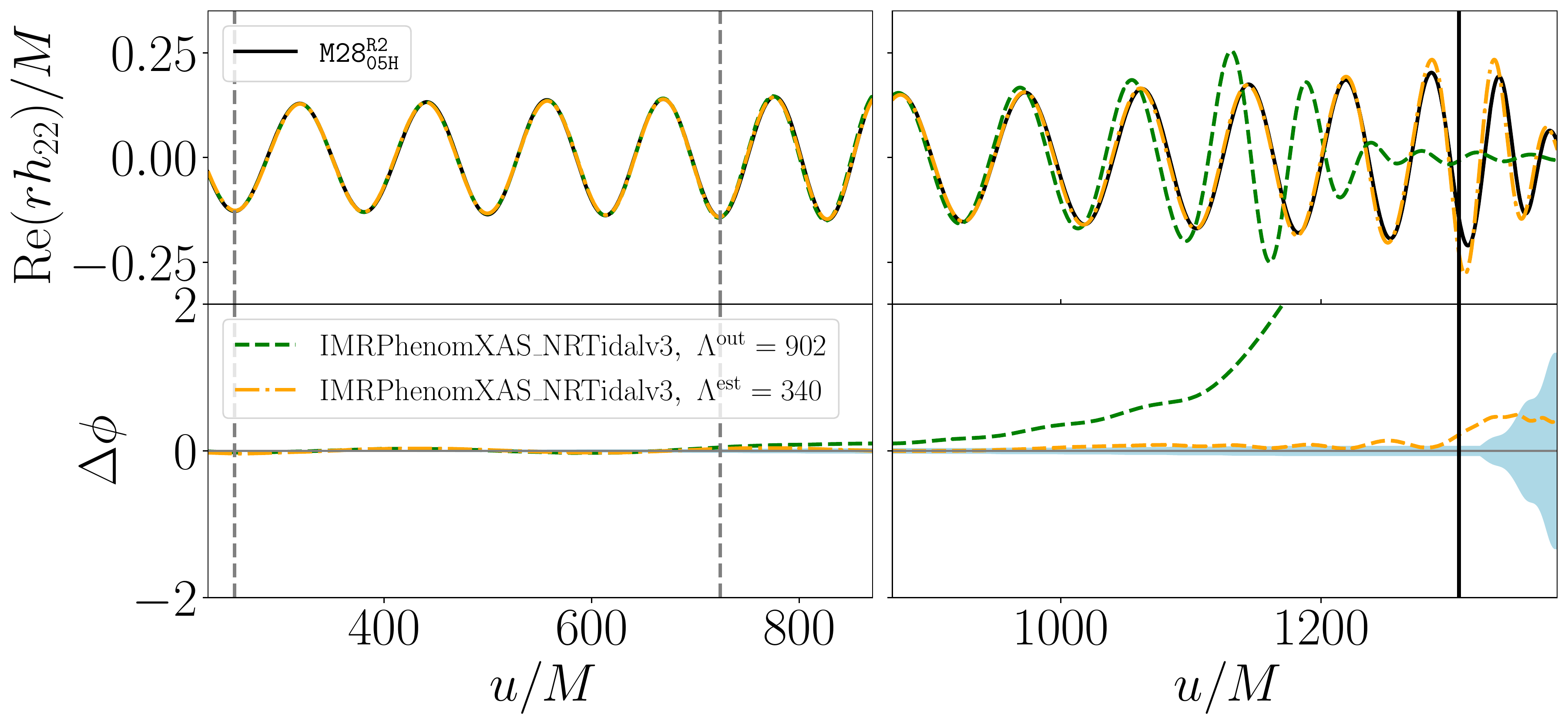}
\caption{Time-domain dephasing comparisons for the different NR configurations
with the \imrphenomxasnrtidalthree{} model. 
For each NR waveform, the upper panel shows the real part of the gravitational
wave strain as a function of the retarded time, while the bottom panel shows the
phase difference between the waveform model and the NR waveform, 
together with the light blue error bands.
The alignment window is indicated by the gray dashed vertical lines in the 
early inspiral, while the merger is indicated by the solid black vertical line.
For the halo configurations, we introduce a comparison with the same waveform approximant but this time with an estimated tidal deformability $\Lambda^{\rm est}$, comparable to those of the other DM configurations. This results in a phase
difference that is also comparable to the phase differences observed for the 
purely baryonic and the DM-core configurations.
The bottom panels (for the DM-halo configurations) show the comparison between 
the two choices for the tidal deformability. 
Green dashed lines represent the GW obtained when using a tidal deformability 
computed assuming that the star extends to the edge of the halo. 
Orange dot-dashed lines show the GW obtained when using an estimated Lambda and 
agree better with the black NR waveform.
}
\label{fig:timedomaincomparisons}
\end{figure*}

We show the results of these comparisons in Fig.~\ref{fig:timedomaincomparisons}. For each NR simulation, we show in the figure the $(2,2)$-mode waveform strain of the highest resolution \texttt{R3} and its phase difference with the model. 
Based on our convergence of the waveforms in
appendix~\ref{AppendixA:GWconvergence}, we take the error band or uncertainty 
as the phase difference between the two highest resolutions of the 
NR simulations. For the configurations without DM and with a DM core, we observe
relatively good agreement between the NR waveform and the phenomenological 
model using $\Lambda^\mathrm{out}$ (see table~\ref{table:IDs} for their values).
However, huge
deviations from the error bands are seen for the halo DM configuration,
using the $\Lambda^{\rm out}$ that was computed under the assumption of a 
two-fluid treatment, where the DM and BM only interact gravitationally, and
without proper consideration of the dilute and extended nature of the halo. 
If we assume the $\Lambda$ to be similar to the other purely baryonic 
configurations the phase difference (orange) is significantly reduced and 
the agreement between the waveform and the model improves.
In Fig.~\ref{fig:timedomaincomparisons} we demonstrate this using estimated
tidal deformabilities $\Lambda^{\rm est} = 810$ and $340$ for the 
\identifier{M24}{~~~}{05H} and \identifier{M28}{~~~}{05H} respectively.
These tidal deformabilities are a factor three smaller than the corresponding
$\Lambda^{\rm out}$. This implies a 
need for more accurate calculations of the tidal deformability for the 
two-fluid system in which one component is much more extended and dilute in
comparison to the second one. Particularly, calculations for DM and BM 
interacting only gravitationally~\cite{Nelson:2018xtr,Rafiei_Karkevandi_2022,
Leung:2022wcf,Giangrandi:2022wht,Liu:2024rix}, 
and where the halo is extended, should be improved.

\begin{figure}[ht!]
    \includegraphics[width=0.95\linewidth]{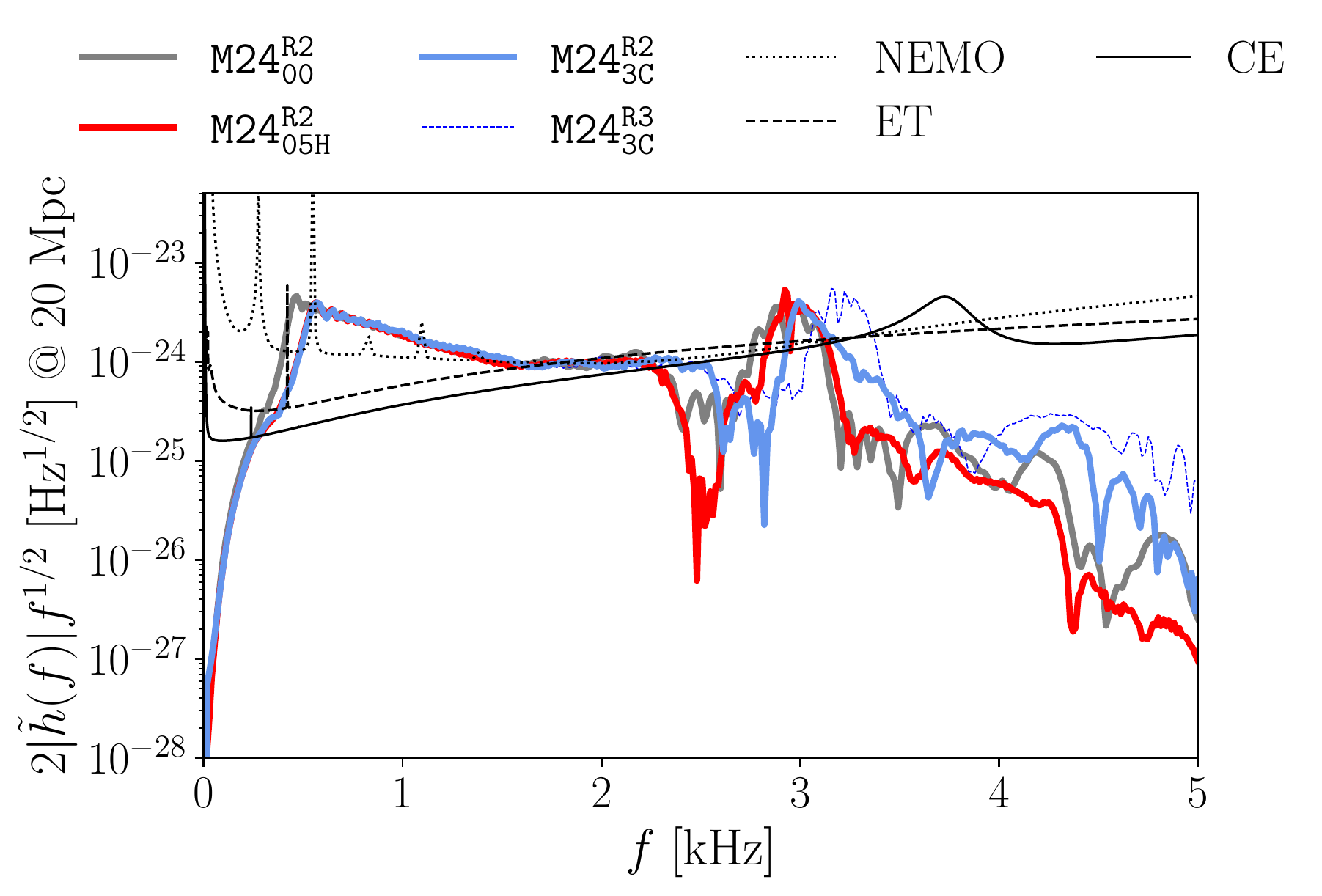}
    \caption{Spectral density of the ${\tt M24}^{\tt R2}$ GW signals extracted at $r_\mathrm{ext}=1474.21$~km as a function of frequency $f$ [kHz].    
    The solid, dashed, and dotted black curves depict the sensitivity curves for Cosmic Explorer (CE), Einstein Telescope (ET), and NEMO respectively.
    }
    \label{fig:PSD-post}
\end{figure}

Figure~\ref{fig:PSD-post} shows the spectral density of the GW signals of
the ${\tt M24}$ simulations as a function of the frequency.  
For frequencies smaller than 0.5~kHz the spectral density exhibits a sharp drop 
in magnitude, which is unphysical and caused by the fact that the inspiral
starts at a finite distance and hence at finite orbital frequency.
Above 0.5~kHz the spectral density exhibits a monotonic exponential decay,
typical for the spectrum of the inspiral phase. At frequencies above 2~kHz 
the spectrum is characterized by the imprint of the post-merger dynamics.
The ${\tt M24}$ simulations end with HMNS remnants,
which allows to test models for the post-merger power spectrum.
An important property is the $f_2$-frequency, which is the frequency of the peak
of the post-merger power spectral density.
The ${\tt M24}^{\tt R2}$ signals, representing simulations with 
the same resolution ${\tt R2}$, show consistent peak frequencies in the
power spectrum of the (2,2)-mode, suggesting
robustness in the dominant mode, $f_{2}^{\rm NR}=2.983,\ 2.923,\ 2.990$ kHz, 
for ${\tt M24}_{\tt 00}^{\tt R2}$, ${\tt M24}_{\tt 05H}^{\tt R2}$, 
${\tt M24}_{\tt 3C}^{\tt R2}$, respectively.
Table~\ref{table:f22peaks} presents the extracted frequencies and their comparison with quasi-universal relation predictions, as detailed in Ref.~\cite{Breschi:2019srl}.
The predictions using $\Lambda^{\rm out}$ deviate from the extracted NR values 
by approximately 10\,\%, except for the halo configuration where the difference 
is approximately 30\,\%. 
Using instead $\Lambda^{\rm est}$ yields a frequency more consistent with that inferred from the NR simulation, with a deviation of $\sim10\,\%$.
This underlines once more the need to improve the calculation of the
tidal deformability of two-fluid systems.

In our highest resolution run, ${\tt M24}_{\tt 3C}^{\tt R3}$, the signal shows 
a noticeable shift towards higher frequencies, \ie{} $f_{2}^{\rm NR}= 3.156$ kHz.
This observed shift suggests a potential a numerical effect arising from the increased resolution of the run.
Disentangling the influence of resolution is crucial for an accurate interpretation of the results. Therefore, further runs are required. Specifically, higher-resolution simulations of the halo and DM-free configurations are essential to 
determine the $f_2$-frequencies more accuratly.

\begin{table}
\begin{tabular}{lccc}
\toprule
identifier       & $f_{2}^{\rm NR}$ [kHz] & $f_{2}^{\rm fit} (\Lambda^{\rm out})$ [kHz] & $f_{2}^{\rm fit} (\Lambda^{\rm est})$ [kHz]\\
\midrule
\identifier{M24}{00}{R2}  & 2.983 & 2.677 & - \\
\identifier{M24}{3C}{R2}  & 2.990 & 2.733 & -\\
\identifier{M24}{3C}{R3}  & 3.156 & 2.733 & -\\
\identifier{M24}{05H}{R2~}& 2.923 & 1.931 & 2.682 \\ \bottomrule
\end{tabular}
\caption{The $f_{2}$ frequencies of the \identifier{M24}{}{} runs compared to the quasi-universal fit of Ref.~\cite{Breschi:2019srl}, evaluated using $\Lambda^{\rm out}$ and the estimated $\Lambda^{\rm est}$.
}
\label{table:f22peaks}\end{table}

\section{Conclusions}\label{section:conclusions}
This study presents, up to our knowledge, the first simulations of DM-admixed BNS systems within a full GR framework employing constraint-solved ID. Quasi-equilibrium configurations are generated using the updated \textsc{sgrid} code~\cite{Ruter:2023uzc}. These ID are constructed employing the SLy4 EOS for BM and modeling DM as a non-interacting Fermi gas. We investigate DM-admixed BNSs with two distinct morphologies: a dense DM core fully embedded within the BM star and a diluted DM halo encompassing the entire BM structure. 
Simulations are performed for total system masses of $2.4 M_\odot$ and 
$2.8 M_\odot$ to better explore the impact of DM on merger dynamics and 
remnant fate in a broader parameter space. We establish a clear comparison 
by running DM-free simulations, allowing us to better isolate DM's effects 
from inherent NR uncertainties. In the following, we summarize our main findings:
\begin{itemize}
    \item {\bf Merger dynamics and DM structure:} 
    The merger time of DM structures is highly sensitive to the spatial 
    extent of the DM component. Extended DM halos come into contact earlier,
    forming a common envelope around the inspiraling BM stars. 
    Contrary, the DM cores remain distinct and tidally 
    deformed when the BM structures come into contact. Notably, the merger
    preserves the DM morphology of the ID. Simulations resulting in 
    an HMNS from DM-core configurations retained a core-like DM structure 
    in their central regions. On the other hand, DM-halo configurations evolve
    into a more diffuse and spatially extended cloud-like DM distribution
    surrounding the HMNS.
    \item {\bf Remnant fate and mass dependence:}
    For lower mass systems, \ie{} \identifier{M24}{}{}, simulations yield HMNS remnants.
    The presence of DM, particularly in core configurations, 
    lead to increased central BM densities and more compact remnants, 
    reflecting the initial conditions of the system where the enhanced
    gravitational pull of the dense DM core creates more tightly bound
    configurations, with greater central densities. 
    On the other hand, higher mass systems, \ie{}, \identifier{M28}{}{}, 
    experience a prompt collapse to a BH. In these cases, DM tends to shorten the
    remnant's lifespan, favoring gravitational instabilities and the 
    formation of a BH. The final BH mass exhibits a weak dependence on the DM
    morphology, with  DM-core simulations resulting in slightly higher BH masses
    compared to DM-free or DM-halo configurations. 
    We emphasize that DM-free simulations may yield lower BH mass due to a 
    longer binary inspiral, potentially leading to greater mass diffusion 
    compared to the other setups.
    \item {\bf Rotational dynamics:} 
    Our analysis of the post-merger rotational evolution suggests potential
    differences in angular velocity based on DM morphology.
    DM cores exhibit higher rotational velocities compared to DM halo
    configurations. While BM angular velocity shows less pronounced changes, 
    halo configurations show the presence of a central region with lower BM
    angular velocities, suggesting a more complex rotational profile.
    \item {\bf BM ejecta masses:} 
    Our simulations reveal the impact of DM morphology on both BM and DM ejecta:
    DM-core configurations can suppress BM ejection in higher mass systems. 
    At the same time, for less massive systems, DM-cores seem to enhance 
    shock-driven BM ejecta via more violent mergers. On the other hand, 
    DM halos generally lead to similar BM ejecta compared to the DM-free 
    scenario. These trends are robust across different resolutions.
    \item {\bf DM ejecta masses:} 
    We observe DM ejecta in the range $[10^{-6}$, $10^{-4}] M_\odot$ in 
    the presented mergers. These trends are robust across different resolutions; however, they are expected to be sensitive on the DM EOS and the temperature dependence of the DM. Furthermore, we expect that the component masses and the BM EOS also play a crucial role in determining the DM ejecta.
    While it is possible that these ejecta may form DM sub-structures and/or
    subsequently be accreted by other stellar objects, this scenario requires
    confirmation through further research. Indeed, the ejected DM during the merger might
    be lately accreted by other objects in the vicinity, leading to
    {\it DM recycling}. Understanding the viability of this mechanism is of
    interest for refining models of DM accumulation in astrophysical objects. 
    \item {\bf GW signal and Tidal Deformability:} 
    GW signal extraction from our simulations reveals a significant deviation 
    in tidal deformability calculations for halo configurations when it is
    calculated in a two-fluid framework. Employing a tidal deformability value similar to that of the pure BM and DM core configurations 
    significantly improves the agreement between the DM-halo configuration and 
    the \imrphenomxasnrtidalthree{} waveforms, highlighting the need for refined
    calculations for DM-admixed NSs. This initial finding indicates that regions
    of the $f_\chi -m_\chi$ parameter space previously considered excluded by the GW data may need to be revisited. This could imply that extended DM halos may indeed be consistent with the GW170817 and GW 190425 binary NS mergers, contrary to conclusions drawn from previous
    results~\cite{Sagun:2021oml}.
\end{itemize}

\begin{acknowledgments}
The authors thank William Newton for providing the SLy4 EOS tabulated to high-density, enabling the exploration of extreme density regimes reached in the simulations. The authors acknowledge Nils Andersson for his significant contributions through discussions and his help in interpreting the analysis.
The work of E.\,G.\@, H.\,R.\,R.\@ and C.\,P.\@ were supported by national funds from FCT – Fundação para a Ciência e a Tecnologia within the projects UIDP/\-04564/\-2020 and UIDB/\-04564/\-2020, respectively, with DOI identifiers 10.54499/UIDP/04564/2020 and 10.54499/UIDB/04564/2020. 
E.\,G.\@ also acknowledges the support from Project No.~PRT/BD/152267/2021. 
E.\,G.\@ and V.\,S.\@ acknowledge the support from FCT – Fundação para a Ciência e a Tecnologia within the project ``Fundamental physics with neutron stars and their mergers'' (Grant No.~2023.10526.CPCA.A2 with DOI identifier 10.54499/2023.10526.CPCA.A2). 
H.\,R.\,R.\@ acknowledges financial support from the FCT – Fundação para a Ciência e a Tecnologia, I.P., within the Project No.~EXPL/FIS-AST/0735/2021. 
H.\,R.\,R.\@ acknowledges financial support provided
under the European Union's H2020 ERC Advanced Grant ``Black holes:
gravitational engines of discovery'' grant agreement no.~Gravitas–101052587. Furthermore, T.\,D.\@ acknowledges funding from the EU Horizon under ERC Starting Grant, no.\ SMArt-101076369. Views and opinions expressed are however those of
the authors only and do not necessarily reflect those of the European
Union or the European Research Council.  Neither the European Union
nor the granting authority can be held responsible for them.
V.\,S.\@ gratefully acknowledges support from the UKRI-funded ``The next-generation gravitational-wave observatory network'' project (Grant No.~ST/Y004248/1).
W.\,T.\@ was supported by NSF grants PHY2136036 and PHY-2408903.
The authors gratefully acknowledge the Gauss Centre for Supercomputing e.V.\ for funding this project by providing computing time on the GCS Supercomputer SuperMUC-NG at Leibniz Supercomputing Centre [project pn29ba], the national supercomputer HPE Apollo Hawk at the High-Performance Computing (HPC) Center Stuttgart (HLRS) under the grant number GWanalysis/44189, and we acknowledge the usage of the DFG-funded research cluster jarvis at the University of Potsdam (INST 336/173-1; project number: 502227537).
\end{acknowledgments}
\section*{Data and Materials availability}
GW data are available on \href{https://doi.org/10.5281/zenodo.16638116}{Zenodo}.
Data supporting the findings of this study are available from the corresponding author on request.
3D videos are available at the following links: \href{https://youtu.be/EVXfDnEgOlI}{\identifier{M28}{3C}{R3}} and \href{https://youtu.be/X27tkz-5aok}{\identifier{M24}{3C}{R3}}.
\twocolumngrid
\appendix
\section{L2-norm of the Hamiltonian constraint}\label{AppendixA:L2norm}
\begin{figure}[t]
    \includegraphics[width=\linewidth]{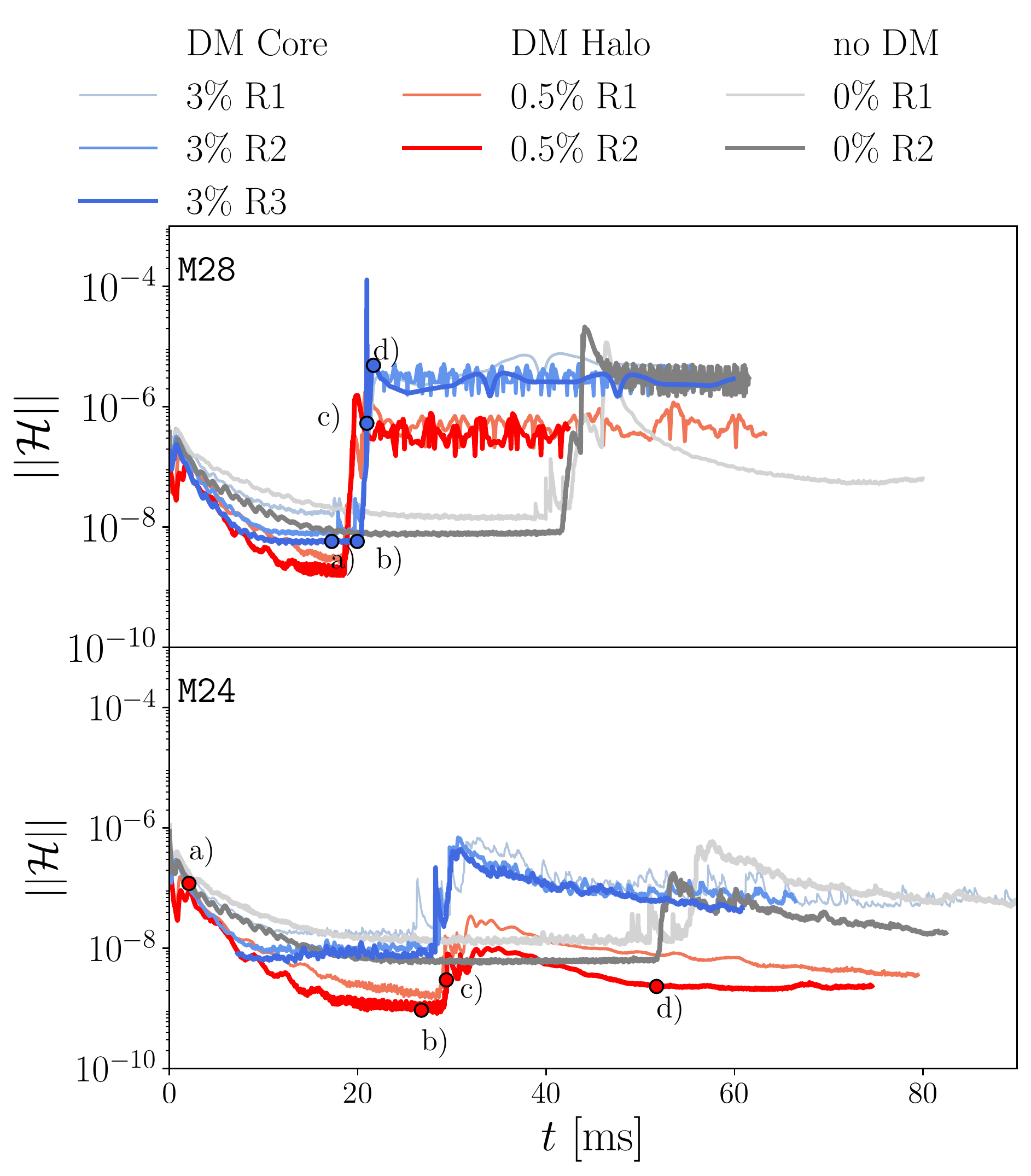}
    \caption{Time evolution of the L2-norm of the Hamiltonian constraints on the
    second to finest refinement level. 
    The top panel shows the values for the \identifier{M28}{}{} setups, 
    while the bottom panel shows the \identifier{M24}{}{} setups.
    }
    \label{fig:ham}
\end{figure}

Figure~\ref{fig:ham} shows the evolution of the L2-norm of the Hamiltonian constraint for the simulations run within this framework. All runs exhibit an initial burst of constraint violation within the first few milliseconds, followed by a rapid decrease related to the constraint damping of the Z4c formalism~\cite{Bernuzzi:2009ex,Hilditch:2012fp}. Following this initial phase, the configurations evolve differently. 
For $2.8 M_\odot$, all runs, except for \identifier{M28}{00}{R1}, 
exhibit a sudden increase in the Hamiltonian constraint violation around 
$t\sim20$ ms. 
Figure~\ref{fig:ham} also shows that \identifier{M28}{00}{R3} has a spike
in $||\mathcal{H}||$ at $t = 20.97 \ {\rm ms}$.  
We note that such a feature is not observed at the lower resolutions. 
This spike is strongly related to the formation of a BH. The collapse of the NS to a BH is a highly dynamic process that generates a strong gravitational field and, consequently, a larger constraint violation. However, we note that 
\identifier{M28}{00}{R1} does not show such a collapse. The post-merger behavior follows the trend expected for a constraint damping formalism. 
On the other hand, the $2.4 M_\odot$ runs, regardless of the DM morphology, 
do not show a sharp increase in violation of the constraints due to the BH
formation. Instead, they show a more `controlled' evolution, although the
different DM profiles affect the magnitude of the constraint violation.

\section{Rest Mass Evolution}

\begin{figure}[t]
    \includegraphics[width=\columnwidth]{Figures/D_t_M24.pdf}
    \caption{Relative change of the rest mass as a function of time for BM (upper panels) and DM (lower panels) for the \identifier{M24}{}{} runs. All the values are extracted at level $l=1$.
    }
    \label{fig:ghrd_rho_integral_l2_M24}
\end{figure}

The performance of the conservative adaptive mesh refinement can be tested by checking the rest mass conservation of the system. The rest mass in the simulations can be evaluated as
\begin{equation}
    M_b^{(s)}=\int d^3x\sqrt{\gamma}D^{(s)}\,.
\end{equation}
This quantity should remain constant during the whole evolution if mass losses (and sources) are not present in the simulation. The rest mass in the \textsc{BAM} code is evaluated on each refinement level.
Figures~\ref{fig:ghrd_rho_integral_l2_M24} and~\ref{fig:ghrd_rho_integral_l2_M28}
show the time evolution of the 
relative change in rest masses $M_b^{(s)}$ evaluated at level $1$ relative to the initial value $M_{b,0}^{(s)}$. 
The figure shows that the relative change in rest masses remains on the order
of $\sim 10^{-5}$ up to the merger, consistent with previous studies~\cite{Dietrich:2015iva}. Lower resolution simulations yield larger variations, indicative of increased numerical errors. 
\begin{figure}[t]
    \includegraphics[width=\columnwidth]{Figures/D_t_M28.pdf}
    \caption{Relative change of the rest mass as a function of time for BM (upper panels) and DM (lower panels) for the \identifier{M28}{}{} runs. All the values are extracted at level $l=1$.
    }
    \label{fig:ghrd_rho_integral_l2_M28}
\end{figure}

Simulations that result in BH formation display a sharp decrease in the rest mass, a known
numerical artifact due to the matter being absorbed by the coordinate puncture
that is created during BH formation. 
Moreover, in simulations without BH formation a noticeable decrease in 
baryonic rest mass is observed
as a significant fraction of ejecta reaches the refinement level boundary 
and exits the computational grid. The same behavior can be seen 
for the DM rest mass in ${\tt M24}_{\tt 05H}$ simulations. In the
${\tt M24}_{\tt 3C}$ simulations on the other hand the amount of DM ejecta is 
too small to significantly change the DM rest mass.

\begin{figure}[t]
    \includegraphics[width=\columnwidth]{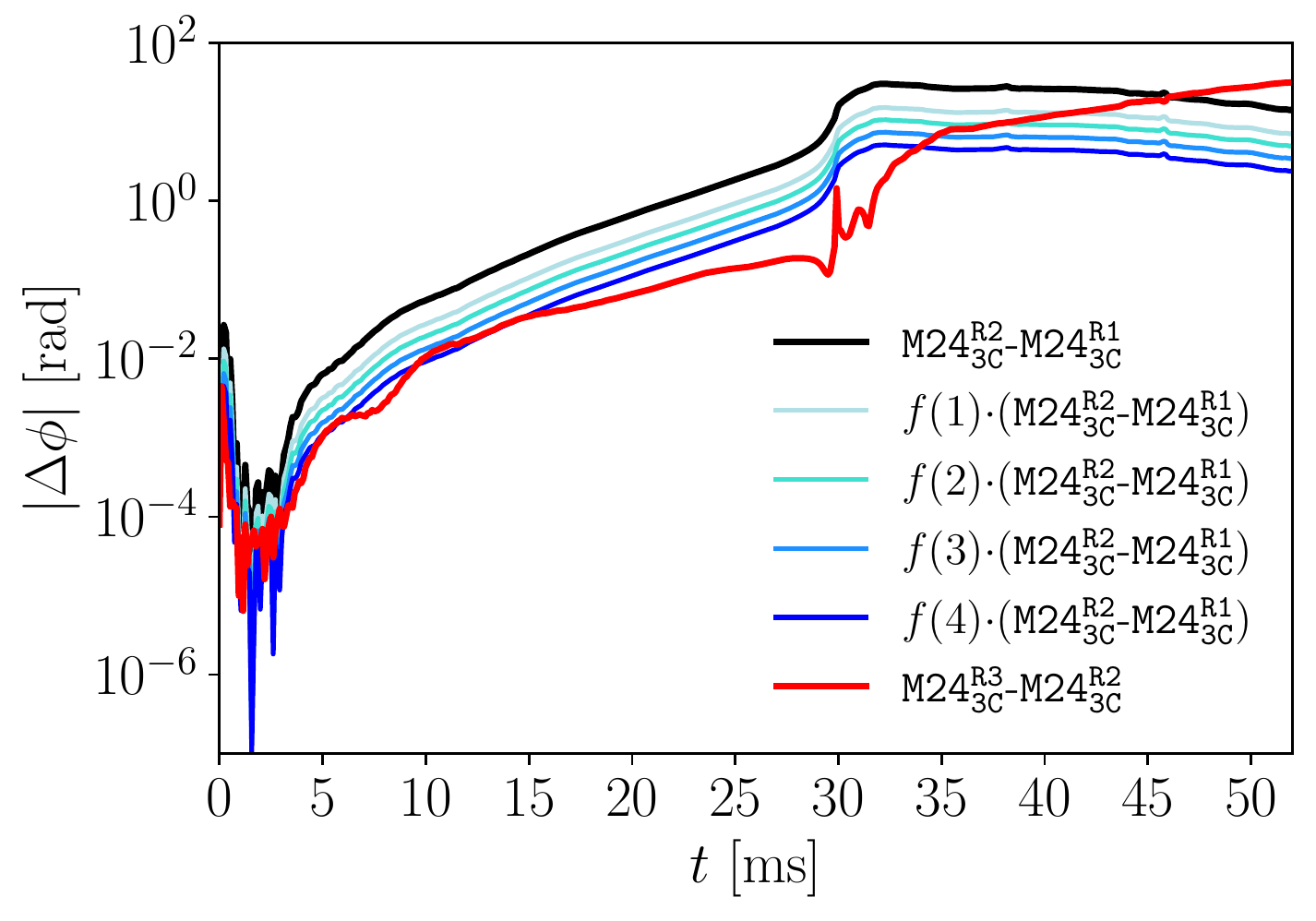}
    \caption{Evolution of the absolute value of the phase difference between the waveforms at different resolutions, 
    \identifier{M24}{3C}{R2}-\identifier{M24}{3C}{R1},
    \identifier{M24}{3C}{R3}-\identifier{M24}{3C}{R2},
    and the rescaled phase difference assuming $n$-th order of convergence.
    }
    \label{fig:convergence_GW_M24}
\end{figure}

\begin{figure}[t]
    \includegraphics[width=\columnwidth]{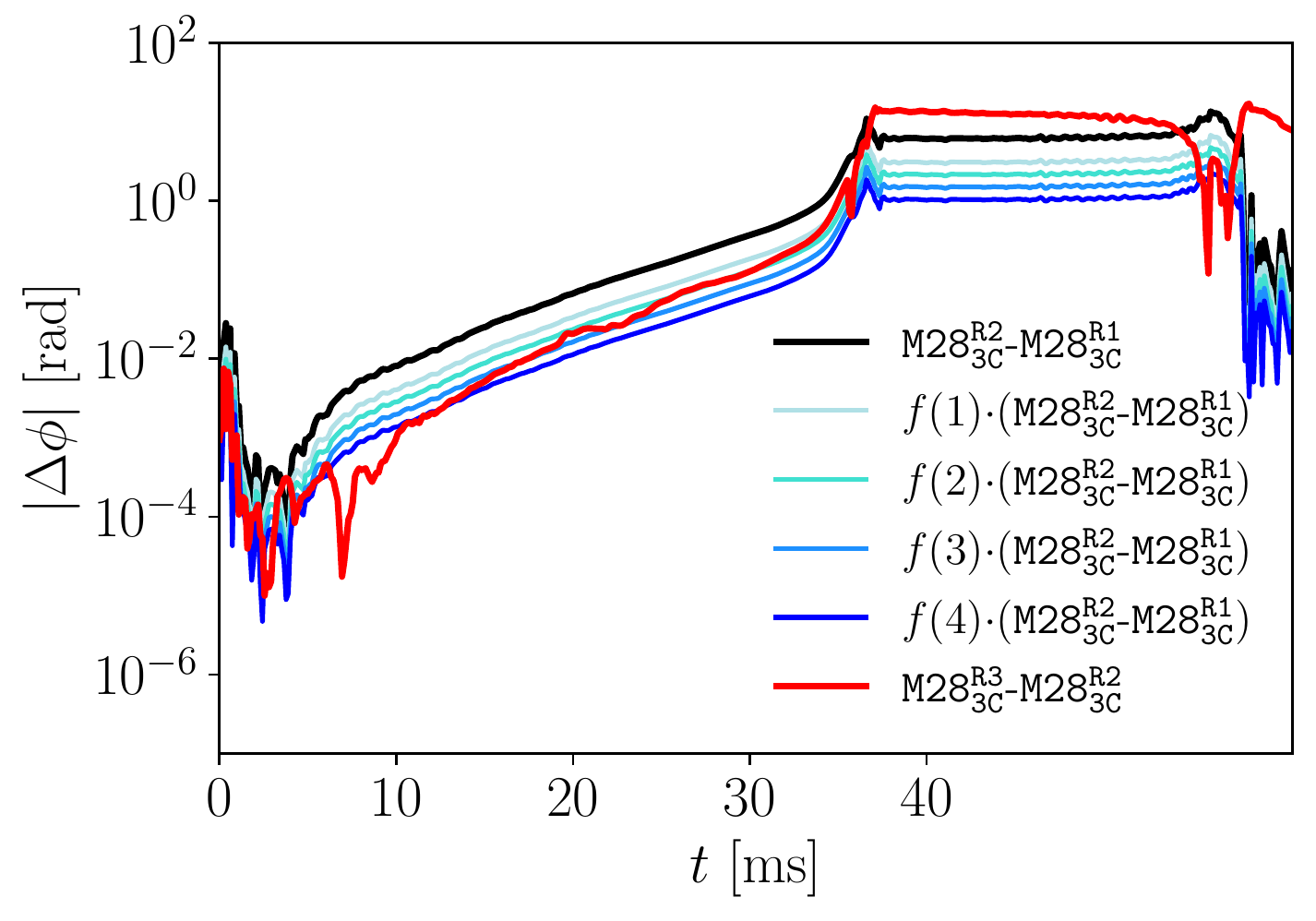}
    \caption{Evolution of the absolute value of the phase difference between the waveforms at different resolutions, 
    \identifier{M28}{3C}{R2}-\identifier{M28}{3C}{R1},
    \identifier{M28}{3C}{R3}-\identifier{M28}{3C}{R2}, 
    and the rescaled phase difference assuming $n$-th order of convergence.
    }
    \label{fig:convergence_GW_M28}
\end{figure}

\section{GW Convergence}\label{AppendixA:GWconvergence}

We test the convergence of waveforms of different resolutions
by comparing the scaled the phase difference between (${\tt R3}$-${\tt R2}$) 
and (${\tt R2}$-${\tt R1}$).
The scaling function used to determine the convergence order is defined as:
\begin{equation}
    f(n) = \frac{\Delta x^n_{\rm R3}-\Delta x^n_{\rm R2}}{\Delta x^n_{\rm R2}-\Delta x^n_{\rm R1}}=\frac{N^{-n}_{\rm R3}-N^{-n}_{\rm R2}}{N^{-n}_{\rm R2}-N^{-n}_{\rm R1}}\,,
\end{equation}
where $\Delta x^n_{\rm Ri}$ and $N^n_{\rm Ri}$ are the grid spacing and the number of points in one dimension in a specific resolution ${\rm Ri}$, respectively, and $n$ is the order of convergence.

As shown in Figs.~\ref{fig:convergence_GW_M24}-\ref{fig:convergence_GW_M28}, there is no clear convergence order that scales the phase difference between the resolutions reasonably during the whole inspiral. 
In the initial part of the waveform the convergence is roughly of fourth order,
consistent with our fourth-order time integration method.
At a later stage shocks start to form in the hydrodynamics part of the
evolution system which lead to a loss of differentiability, which is known
to limit the achievable order of convergence. In particular around the
merger this typically causes a more erratic convergence behaviour as also
observed in Figs.~\ref{fig:convergence_GW_M24} and~\ref{fig:convergence_GW_M28}.
Regardless of this, up to merger the phase difference decreases with increasing
resolution in a monotonic and consistent manner. Given the lack of convergence
order, we adopt the difference between the highest resolution as an error 
estimate in Figs.~\ref{fig:timedomaincomparisons}.
\section{\texorpdfstring{$(2,|1|)$}{(2,|1|)}-GW mode}\label{Appendix:21mode}
\begin{figure}[t]
    \centering
    \includegraphics[width=0.95\linewidth]{Figures/PSD_post_l2m1.pdf}
    \caption{Spectral density of the $l=2, |m|=1$ mode for \identifier{M24}{}{} GW signals extracted at $r_\mathrm{ext}=1474.21$~km as a function of frequency $f$ [kHz].
    The GW signal is scaled for a distance of $20~{\rm Mpc}$ to the source.
    The solid, dashed, and dotted black curves depict the sensitivity curves for Cosmic Explorer (CE), Einstein Telescope (ET), and NEMO respectively.}
    \label{fig:PSD_post_l2m1}
\end{figure}
Figure~\ref{fig:PSD_post_l2m1} presents the characteristic spectral density of
the $l=2, |m|=1$ mode for the ${\tt M24}^{\tt R2}$ and 
${\tt M24}^{\tt R3}$ configurations. 
This mode becomes prominent in compact binary merger events characterized by precessing orbital planes or significant mass asymmetries. 
As the NSs in our configurations have equal masses and no spin, this mode
should have vanishing amplitude. However, numerical round-off introduces
small deviations from this symmetry, causing a small excitation of the
mode. \citet{Bezares:2019jcb} observed a similar behavior and argued 
that the presence of DM could enhance the growth of this asymmetry.

The peaks of the $l=2, |m|=1$ mode are observed at $f^{\rm peak}_{(l=2, |m|=1)}=1.512, 1.461, 1.541~{\rm kHz}$ for \identifier{M24}{00}{R2}, \identifier{M24}{3C}{R2} and \identifier{M24}{05H}{R2}, respectively. 

Notably, DM-core configurations exhibit a second peak that is located at 
roughly twice the frequency of the first one. The additional peak is observed at
$f^{\rm extra}_{(l=2, |m|=1)}=2.991,3.156~{\rm kHz}$ for \identifier{M24}{3C}{R2} and \identifier{M24}{3C}{R3}, respectively. This contribution is notably absent in DM-halo configurations, and is not as evident in the DM-free runs, strongly indicating that its presence may depend on the specific properties and morphology of DM. Specifically, the presence of these DM cores enhances asymmetries, initially caused by numerical roundoff, leading to an increase of the amplitude of this particular GW mode.
\vfill
\bibliography{main.bbl}
\end{document}